\documentclass[5p,longtitle, number, sort&compress]{elsarticle}

\pdfoutput=1
\usepackage[utf8]{inputenc}
\usepackage{lineno}
\usepackage{import}
\usepackage{siunitx}
\usepackage{amssymb}
\usepackage{multirow}
\usepackage{array}
\usepackage[italic]{heppennames}
\usepackage{footnote}

\usepackage[dvipsnames]{xcolor} 

\RequirePackage{xspace}

\RequirePackage{hyperref}
\begin{document}

\begin{frontmatter}

\title{The Data Acquisition System of the LZ Dark Matter Detector: FADR}

\author[1,2]{J.~Aalbers}
\author[1,2]{D.S.~Akerib}
\author[3]{A.K.~Al Musalhi}
\author[3]{F.~Alder}
\author[4,5]{C.S.~Amarasinghe}
\author[1,2]{A.~Ames}
\author[1,2]{T.J.~Anderson}
\author[6]{N.~Angelides}
\author[6]{H.M.~Ara\'{u}jo}
\author[7]{J.E.~Armstrong}
% ...

\address[1]{SLAC National Accelerator Laboratory, Menlo Park, CA 94025-7015, USA}

\address[2]{Kavli Institute for Particle Astrophysics and Cosmology, Stanford University, Stanford, CA  94305-4085 USA}

\address[3]{University College London (UCL), Department of Physics and Astronomy, London WC1E 6BT, UK}

\address[4]{University of California, Santa Barbara, Department of Physics, Santa Barbara, CA 93106-9530, USA}

\address[5]{University of Michigan, Randall Laboratory of Physics, Ann Arbor, MI 48109-1040, USA}

\address[6]{Imperial College London, Physics Department, Blackett Laboratory, London SW7 2AZ, UK}

\address[7]{University of Maryland, Department of Physics, College Park, MD 20742-4111, USA}

% ...


\begin{abstract}
The Data Acquisition System (DAQ) for the LUX-ZEPLIN (LZ) dark matter detector is described.  The signals from 745 PMTs, distributed across three subsystems, are sampled with 100-MHz 32-channel digitizers (DDC-32s).  A basic waveform analysis is carried out on the on-board Field Programmable Gate Arrays (FPGAs) to extract information about the observed scintillation and electroluminescence signals.  This information is used to determine if the digitized waveforms should be preserved for offline analysis.

The system is designed around the Kintex-7 FPGA. In addition to digitizing the PMT signals and providing basic event selection in real time, the flexibility provided by the use of FPGAs allows us to monitor the performance of the detector and the DAQ in parallel to normal data acquisition.  

The hardware and software/firmware of this FPGA-based Architecture for Data acquisition and Realtime monitoring (FADR) are discussed and performance measurements are described.

\end{abstract}


\end{frontmatter}


\section{Introduction}\label{sec:Intro}

\begin{figure}
	\centering
	\includegraphics[width=1.0\linewidth]{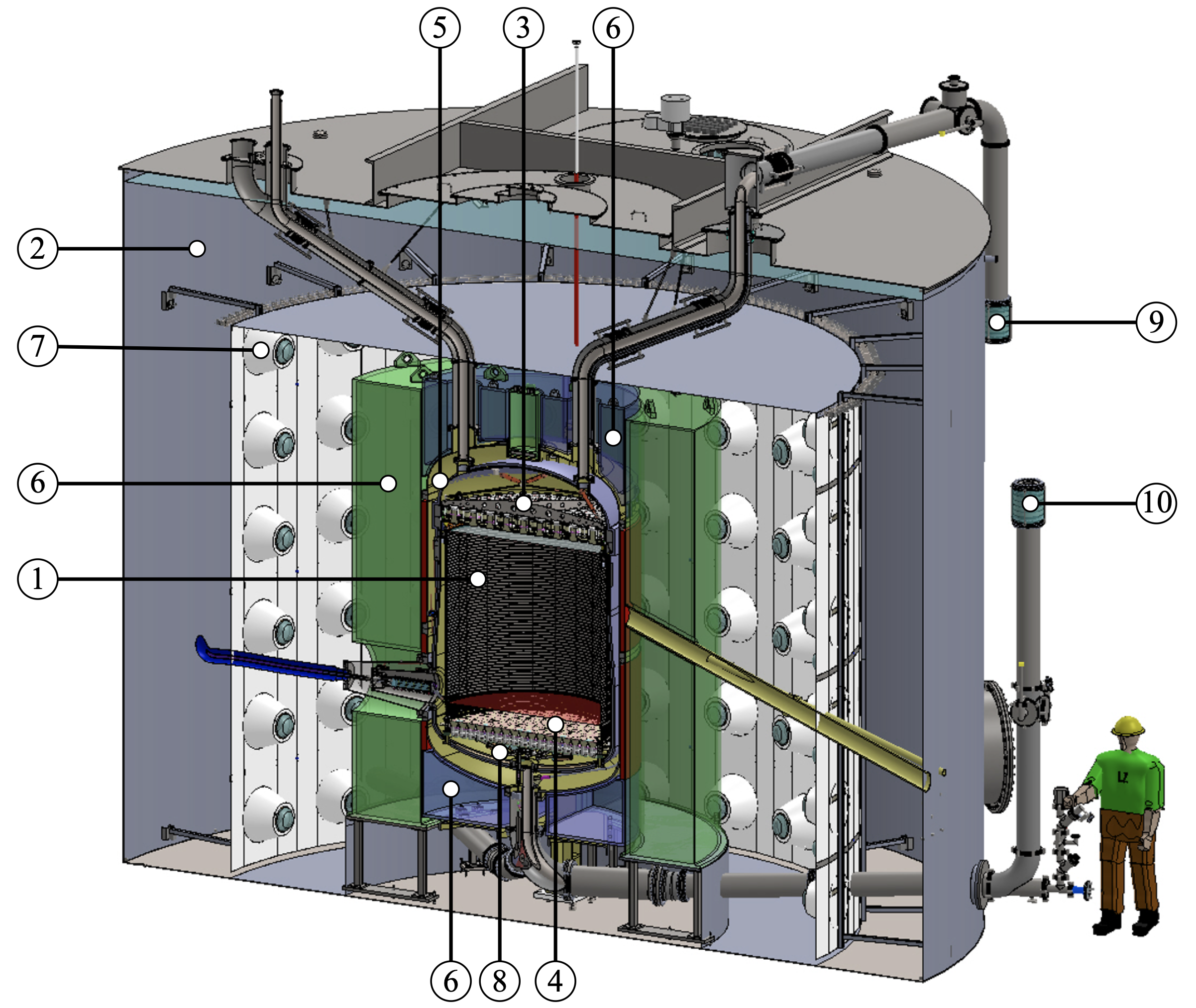}
	\caption{Schematic of the LZ detector with its main components indicated.  The liquid xenon time-projection chamber (1), located in the center of a water tank (2), is monitored by two arrays of photomultiplier tubes (PMTs), the top TPC PMT array (3) and the bottom TPC PMT array (4). The TPC is contained in a double-walled vacuum insulated titanium cryostat (5) and surrounded on all sides by a gadolinium-loaded liquid scintillator (6). Scintillation light emitted when background radiation interacts with the liquid scintillator is observed with an array of PMTs (7), mounted inside the water tank. The Skin PMTs are installed in the Dome (8) and around the vertical section of the TPC.  The signal and HV cables from the TPC and Skin PMTs are routed through the top (9) and the bottom (10) conduits to the breakout area where the analog electronics are installed.}
	\label{fig:3DModel}
\end{figure}

LUX-ZEPLIN (LZ) is the G2 Dark Matter Search experiment deployed at the Sanford Underground Research Facility (SURF) as a successor to the Large Underground Xenon (LUX) detector \cite{Akerib:2015ega}. Figure~\ref{fig:3DModel} depicts the main parts of the experiment.  The center of LZ is a dual-phase xenon Time Projection Chamber (TPC), surrounded by a Skin detector. The cryostat with the TPC and the Skin detector is surrounded by an Outer Detector (OD) of gadolinium loaded scintillator (GdLS).  The OD is used to tag neutron-scattering events within the TPC.  The OD is installed inside a large water tank to shield the TPC from external gamma rays and neutrons.  More details about the LZ detector components can be found in Ref.~\cite{Akerib:2019fml}.

\begin{figure}
	\centering
	\includegraphics[width=1.0\linewidth]{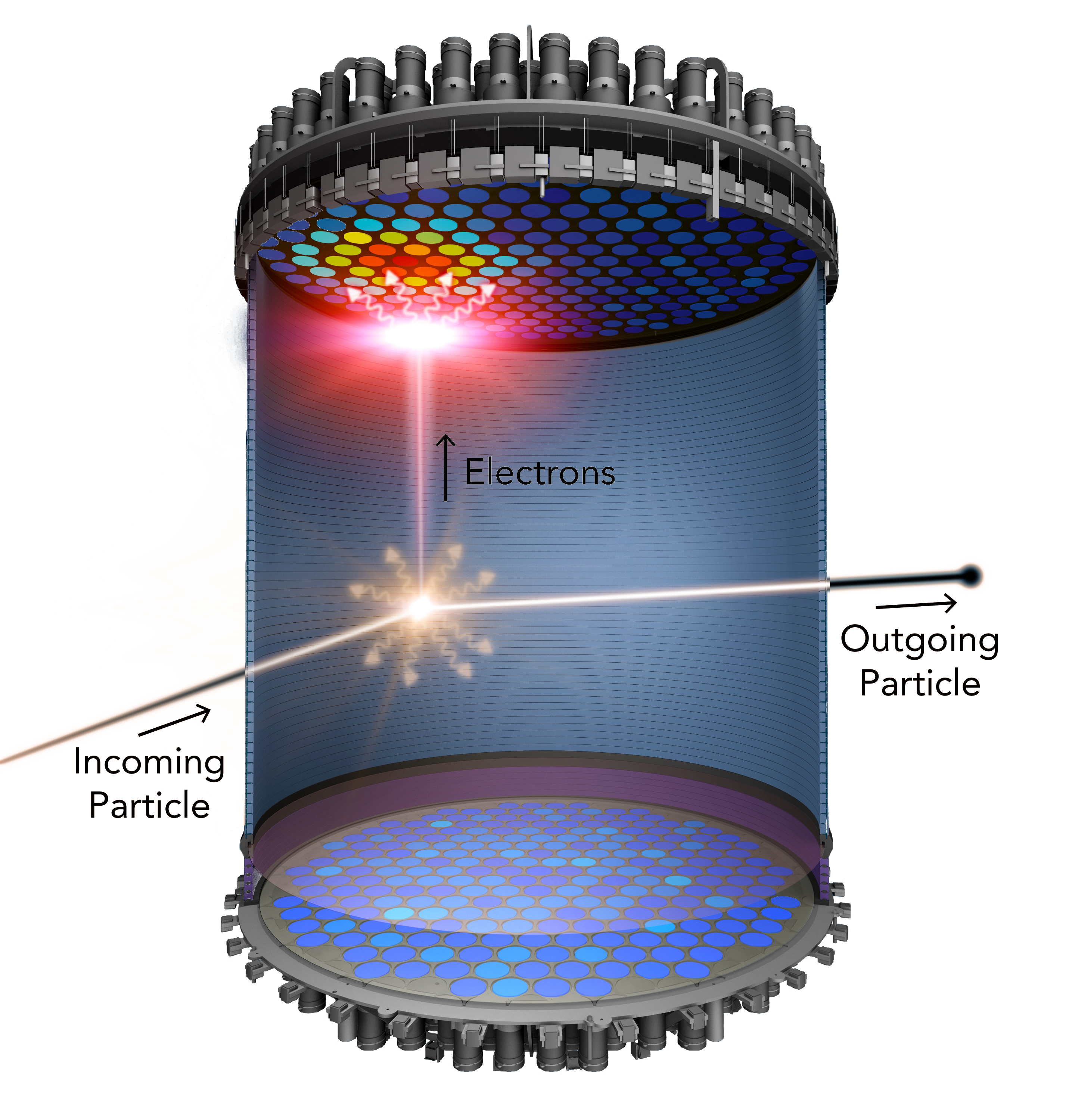}
	\caption{Principle of operation of a dual-phase xenon detector, showing the initial interaction, producing prompt scintillation photons (S1) and ionization electrons that drift towards the surface of the liquid xenon where they produce an electroluminescence signal (S2). A typical S2 pattern is shown on the top PMT array.}
	\label{fig:principleOfOperation}
\end{figure}

The principle of operation of the TPC is shown in Fig.~\ref{fig:principleOfOperation}. Scattering events in the liquid xenon (LXe) create both a prompt scintillation signal (S1) and ionization electrons. The S1 signals typically have a full width at half maximum (FWHM) of less than 100~ns.  Electric fields are employed to drift these electrons to the liquid surface, extract them into the gas phase, and accelerate them to create a proportional scintillation signal (S2) that occurs up to $700 \mu s$ after the S1 and typically have widths of several micro seconds (FWHM). The electric fields are created by various grids at the top and bottom of the xenon volume and a PTFE-clad field cage.  The S1 and S2 signals are detected by arrays of 3-inch diameter photomultiplier tubes (PMTs), 253 of those in the gas above the LXe (top array) and another 241 immersed in the liquid below the active LXe volume (bottom array).  The profile of the S2 light distribution on the top array provides information on the horizontal position of the point of interaction.  The time difference between the S1 and the S2 signals provides information about the depth of the interaction.

The xenon Skin layer includes the xenon between the field cage and the inner wall of the cryostat as well as the region beneath the bottom PMT array.  There are 93 1-inch $\times$ 1-inch PMTs located outside the top of the field cage, looking down towards the xenon Skin layer and 20 2-inch diameter PMTs looking up at the same region. In addition, there are 18 2-inch diameter PMTs located in the Dome region, below the bottom PMT array. Events in the Skin produce S1-like signals.

Interactions in the OD are observed with 120 8-inch diameter PMTs, mounted inside the water tank on stainless steel ladders.  Events in the OD produce S1-like signals.

The Skin and the OD suppress backgrounds in the TPC by tagging $\gamma$-rays and neutrons correlated with dark matter candidates.  The Skin is tagging prompt $\gamma$-ray coincidences with the S1 signals in the TPC.  The OD also sees prompt coincidences but the typical capture times of neutrons in the GdLS and the water are $30~\mu s$ and $220~\mu s$, respectively Ref.~\cite{Akerib:2019fml}.   

This paper describes the data acquisition system used to digitize the signals from the PMTs and identify events of interest for offline analysis.  The system described is an FPGA-based Architecture for Data acquisition and Realtime monitoring (FADR). The system was designed based on our experience with the data acquisition and trigger systems of the Large Underground Xenon (LUX) detector \cite{akerib:2011ix,Akerib_2016}.  Instead of using separate systems for making trigger decisions and acquiring and building events, as was done in LUX, FADR uses the same digitized waveforms for both functions. Since the number of channels to be digitized in LZ is significantly larger than the number of channels used in LUX, reducing the per-channel duplication resulted in significant cost savings. Since we are in full control of the firmware on the FPGAs that processes the digitized waveforms, we can implement novel acquisition modes and detector monitoring tools that would be difficult to implement on commercial digitizers. 

This paper is organized as follows. Section~\ref{sec:Signals} describes the processing of the analog signals before digitization.  The hardware of FADR is described in Sec.~\ref{sec:digitalHardware}.  The design of FADR is presented in Sec.~\ref{sec:DAQ} and the firmware developed for FADR is discussed in Sec.~\ref{sec:firmware}.  Section \ref{sec:eventSelection} describes the process of event selection based on information from various trigger sources. Typical event sizes obtained during normal operation and during calibrations are summarized in Sec.~\ref{sec:eventSize}. Extended functionality, implemented in firmware, is discussed in Sec.~\ref{sec:extraFunctions}.  The performance of FADR is presented in Sec.~\ref{sec:Performance} and the paper is summarized in Sec.~\ref{sec:Summary}.  Abbreviations used in this paper are listed in Table~\ref{tbl:abreviations}.

\begin{table}
\caption{Abbreviations used in this paper.}
\label{tbl:abreviations}
\centering
\vspace{0.25cm}
\small
\begin{tabular}{ p{0.25\linewidth} p{0.65\linewidth} } 
    \hline
    Abbreviation & Description \\
    \hline
    ADC & Analog-to-Digital Converter \\
    DAC & Digital-to-Analog Converter \\
    ADCC & ADC Count (0.122 mV for a 14-bit 2-V ADC) \\
    DAQ & Data Acquisition System \\
    DD & Neutron calibration source that uses deuterium fusion to generate neutrons \\
    DDC-32 & 32-channel Digitizer \\
    FADR & FPGA-based Architecture for Data acquisition and Realtime monitoring \\
    FIR & Finite-Impulse-Response \\
    FPGA & Field Programmable Gate Array \\
    LEMO{\textregistered} & Type of push-pull self-latching connectors, Registered trademark of LEMO, SA \\
    LMR{\textregistered} & Type of low-loss coaxial cables, Registered trademark of Times Microwave Systems \\
    LVDS & Low-Voltage Differential Signaling \\
    LVTTL & Low-Voltage Transistor Transistor Logic \\
    NIM & Nuclear Instrumentation Module \\
    LXe & Liquid Xenon \\
    OD & Outer Detector \\
    POD & Pulse Only Digitization \\
    PPS & Pulse Per Second \\
    RAID & Redundant Array of Independent Disks \\
    SPHE & Single Photoelectron \\
    SR1 & LZ's first Science Run \\
    VHDL & Very High-Speed Integrated Circuit Hardware Description Language \\
    VME & Versa Module Europe, one of the early open-standard backplane architectures \\
    WIMP & Weakly Interacting Massive Particle \\
    \hline
\end{tabular}
\end{table}

\section{The Signal Chain}\label{sec:Signals}

\begin{figure*}
	\centering
	\includegraphics[width=1.00\linewidth]{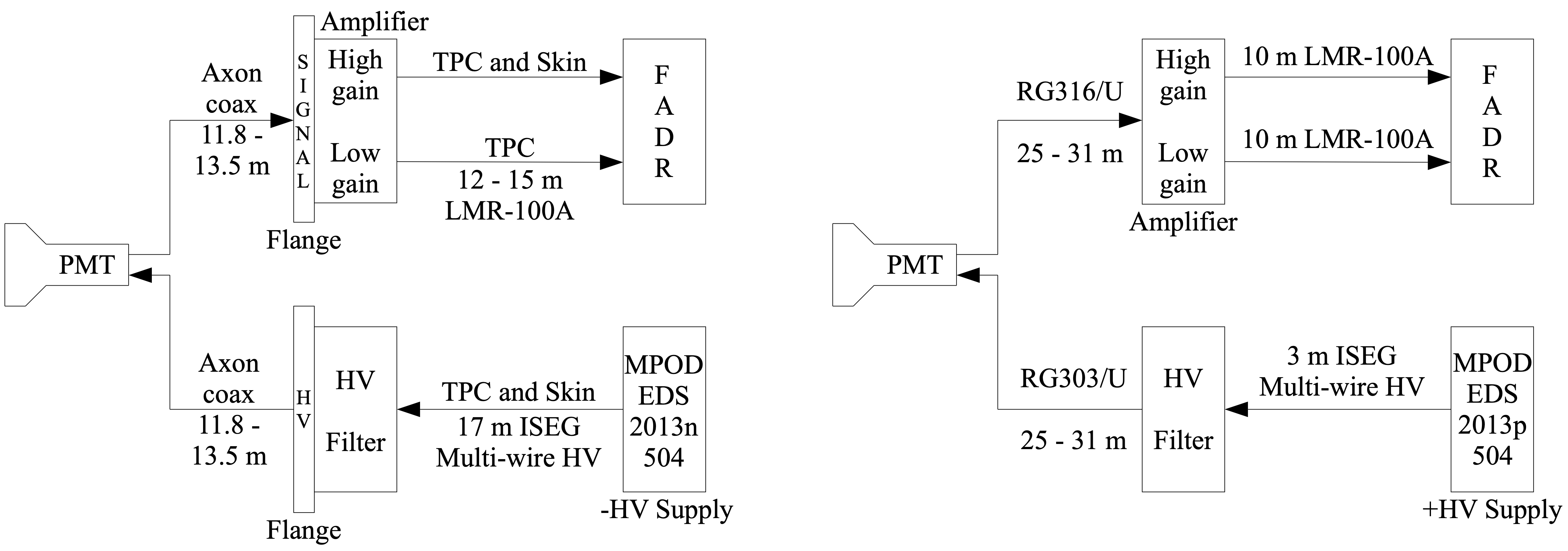}
	\caption{Left: signal chain of the PMTs installed in the xenon space.  The xenon PMTs operate at negative high-voltage (HV) supplied by MPOD EDS 20130n 504 distribution modules from WIENER Power Electronics \cite{WIENER}.  HV filters are mounted onto HV flanges.  The PMT signals are connected to dual-gain amplifiers that are mounted onto signal flanges.  The output signals of the amplifiers are digitized by FADR. The coaxial cables in the xenon space were made by Axon Cable S.A.S. \cite{AxonCable}. The LMR cables between the amplfiers and FADR were produced by Times Microwave Systems \cite{LmrCable}. The signal and HV flanges were made by Accu-Glass \cite{AccuGlass}.  Right: signal chain of the OD PMTs. The signal chain of the OD PMTs does not involve signal and HV flanges. The PMT signal and HV cables are an integral part of the OD PMT assemblies. The cables leave the water tank through a cable port on top of the water tank and connect directly to the amplifiers and HV supplies. The OD PMTs operate at positive HV.}
	\label{fig:signalChain}
\end{figure*}

\begin{figure*}
	\centering
	\includegraphics[width=1.0\linewidth]{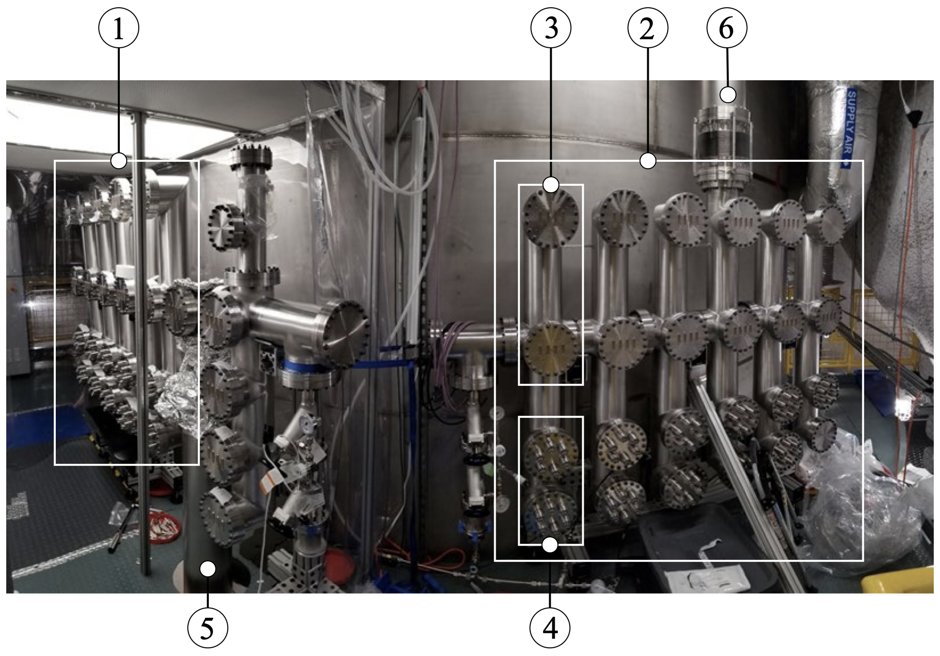}
	\caption{Panoramic image of the two breakout boxes used to provide transitions from xenon to air for the signal and HV cables.  The breakout box on the left-hand side (1) connects to the bottom TPC and Skin PMTs; the breakout box on the right-hand side (2) connects to the top TPC and Skin PMTs.  The top half of each vertical section (3) connects the signal lines; the bottom half of each vertical section (4) connects the high-voltage lines. Also visible are the cable conduits for the bottom (5) and top (6) xenon-space cables. This image was taken before the installation of the analog electronics and the external signal and HV cables.}
	\label{fig:BreakoutBoxes}
\end{figure*}

Block diagrams of the PMT signal chains for the xenon and the OD PMTs are shown in Fig.~\ref{fig:signalChain}. The PMT signals from within the cryostat are routed via thin coaxial cables to custom-made vacuum flanges with four DB-25 feedthroughs each. These flanges are mounted on breakout boxes located on the side of the water tank, shown in Fig.~\ref{fig:BreakoutBoxes}. A set of four 8-channel dual-gain amplifiers are mounted on each signal flange and connect to the PMT signal lines using the DB-25 connectors. The amplifiers are housed in 5-card mini crates that mount directly onto the signal flanges. The fifth card in each crate is a control card used for power distribution and slow-control monitoring. The outputs of the amplifiers are routed to FADR, which is installed in three dedicated electronics racks.  These three electronics racks contain the digital electronics used to digitize all PMT signals (TPC, Skin, and OD).  The OD PMT signal cables are routed to the OD amplifiers located in a dedicated electronics rack.  The outputs of the OD amplifiers are routed to FADR.

The TPC and Skin PMTs operate at a negative high-voltage (HV). The HV filters for the TPC and Skin PMTs are connected to custom-made HV vacuum flanges, installed on the breakout boxes.  Each HV flange has 7 HV feedthroughs; each feedthrough supports up to 6 HV connections.  The OD PMTs operate at a positive HV.  The OD HV filters are installed in the electronics rack for the OD, next to the HV supplies.

The total number of PMT channels processed by FADR is summarized in Table~\ref{tbl:ChannelCounts}. The signal properties for single photoelectrons (SPHEs) at the input of the digitizers are shown in Table~\ref{tbl:SPHEatADC}.  The numbers quoted are for signals processed by the high-gain channels of the amplifiers and include the effects of cable attenuation.  For comparison, the total noise in the system is less than 4~ADC Count (1$\sigma$) for all PMT channels. Note that 1 ADC Count (ADCC) corresponds to 0.122 mV for a 14-bit 2-V ADC.

\begin{table}
\caption{Number of PMTs and channels.  All PMT signals are processed with high-gain amplifiers with an area gain (the increase in the pulse area of the output pulse) of 40 and a Gaussian shaping time of 60 ns full-width at one-tenth maximum (FWTM).  The TPC and OD PMT signals are also processed with low-gain amplifiers with an area gain of 4 and a shaping time of 30 ns (FWTM). All PMTs were manufactured by Hamamatsu~\cite{HAMAMATSU}.}
\label{tbl:ChannelCounts}
\centering
\vspace{0.25cm}
\small
\begin{tabular}{ >{\centering\arraybackslash} p{0.15\linewidth} >{\centering\arraybackslash} p{0.15\linewidth} >{\centering\arraybackslash} p{0.10\linewidth} >{\centering\arraybackslash} p{0.10\linewidth} >{\centering\arraybackslash} p{0.10\linewidth} >{\centering\arraybackslash} p{0.10\linewidth} }
    \hline
    Detector Volume & PMT Model & \# of PMTs & High Gain & Low Gain & Total Channels\\
    \hline
    TPC & R11410-20 & 494 & 494 & 494 & 988 \\
    Top Skin & R8520 & 93 & 93 & 0 & 93 \\
    Bottom Skin & R8778 & 20 & 20 & 0 & 20 \\
    Dome Skin & R8778 & 18 & 18 & 0 & 18 \\
    OD & R5912 & 120 & 120 & 120 & 240 \\
    \hline
    Total & & 745 & 745 & 614 & 1359 \\
    \hline
\end{tabular}
\end{table}

\begin{table}
\caption{The response to SPHE signals from the TPC, Top and Bottom Skin, Dome Skin, and OD PMTs, processed with the high-gain channels of the amplifier, at the input of the digitizer.  The values reported here are average values and their standard deviations of all SPHE PMT signals obtained for each detector volume during LED calibrations.  The TPC, Bottom Skin, Dome, and OD PMTs are operating at a gain of 2$\times 10^6$ but use different values for the load resistor in the PMT base. The Top Skin PMTs are operating at a gain of~1$\times 10^6$.}
\label{tbl:SPHEatADC}
\centering
\vspace{0.25cm}
\small
\begin{tabular}{ >{\centering\arraybackslash} p{0.2\linewidth} >{\centering\arraybackslash} p{0.2\linewidth} >{\centering\arraybackslash} p{0.2\linewidth} >{\centering\arraybackslash} p{0.2\linewidth} }
    \hline
    Detector Volume & Amplitude (mV) & Amplitude (ADCC) & Area (ADCC sample) \\
    \hline
    TPC & 5.9 $\pm$ 0.4  & 49 $\pm$ 3 & 170 $\pm$ 12  \\
    Top Skin & 6.1 $\pm$ 0.7 & 50 $\pm$ 5 & 180 $\pm$ 20 \\
    Bottom Skin & 6.1 $\pm$ 0.3 & 50 $\pm$ 3 & 175 $\pm$ 10  \\
    Dome Skin & 5.5 $\pm$ 0.6 & 45 $\pm$ 5 & 160 $\pm$ 20  \\ 
    OD & 14 $\pm$ 1 & 120 $\pm$ 5 & 415 $\pm$ 15 \\
    \hline
\end{tabular}
\end{table}

\section{Digital Hardware}\label{sec:digitalHardware}

\begin{figure*}
	\centering
 		\includegraphics[width=1.0\linewidth]{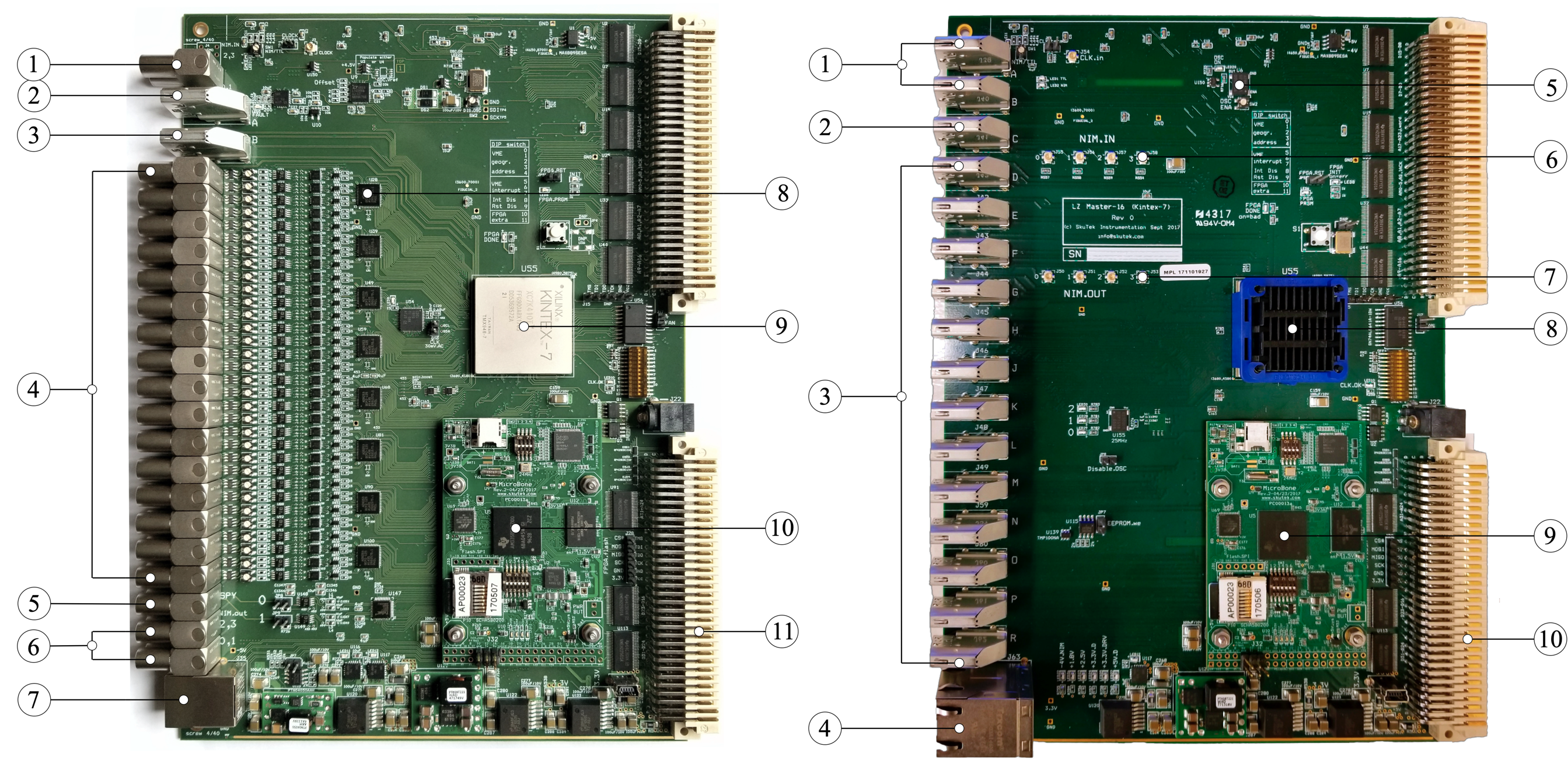}
		\caption{Left: The 32-channel digitizer (DDC-32). (1)~Logic NIM/TTL inputs, (2)~HDMI connection to Data Extractor, (3)~HDMI connection to Data Sparsifier, (4)~32 analog LEMO inputs, (5)~two spy LEMO outputs, (6)~4 NIM outputs, (7)~gigabit Ethernet connection, (8)~four-channel AD9253 ADCs, (9)~KINTEX-7 FPGA before its heatsink was installed, (10)~Arm Cortex-8 (AM3358 processor), and (11)~VME connectors used for power. Right: The logic board used to extract data from the DDC-32s and to make event selections. (1) + (3): 15 HDMI links to up to 15 digitizers, (2)~HDMI link to higher level event selection, (4)~stacked ethernet connectors: one FPGA driven link for data offloading and one MicroBone link for setup/control, (5)~oscillator, (6)~4 NIM MMCX inputs, (7)~4 NIM MMCX outputs, (8)~KINTEX-7 FPGA with heatsink, (9)~AM3358 processor, and (10)~VME connectors for power.}
		\label{fig:daqHardware}
\end{figure*}

FADR uses 47 32-channel 100-MHz digitizers (DDC-32) and 24 logic boards.  Of these 47 DDC-32s, 45 digitize the PMT waveforms and one digitizes the waveforms from the fast sensors, such as loop antennas and acoustic sensors, that are used to monitor the electrostatic environment of the detector \cite{Akerib:2019fml}.  One additional DDC-32 is used to provide the analog signal used for the DAQScope, discussed in detail in Section~\ref{subsec:daqScope}. The digitizers and logic boards, shown in Fig.~\ref{fig:daqHardware}, were developed by Skutek \cite{SkuTek} and are based on the Series-7 Kintex FPGA from Xilinx ~\cite{Xilinx:2016}. The boards provide Gigabit Ethernet, RS-232, and Low-Voltage Differential Signaling (LVDS) interfaces, and four logic outputs, which can be changed to deliver either Low-Voltage Transistor Transistor Logic (LVTTL) or NIM signals. Waveform memory (3578~kB) and very high processing power are provided by the FPGA. The onboard clock can be driven externally in order to synchronize multiple boards to the same clock source. An Arm Cortex-8 (AM3358) processor \cite{AM3358} is connected to the FPGA with a 16-bit wide General Purpose Memory Controller (GPMC). The processor is installed on a MicroBone Single Board Computer (SBC), developed by SkuTek \cite{SkuTek}, which provides 512 MB of DDR3 RAM. The processor is running Linux, allowing each digitizer to perform online diagnostics.  Diagnostic readout of the FPGA data can be performed via the AM3358 GPMC memory bus. Power consumption is minimized by using low-voltage chips. The DDC-32s are installed in 6U VME crates.

Each DDC-32 has two 14-bit, 100-MHz Analog-to-Digital Converter (DAC) "spy" outputs, allowing for diagnostics and monitoring. Individual incoming channels or their digital sum can be sent to the spy outputs. Digital-filter outputs, individual or summed, or any other signal internal to the FPGA can also be viewed using the spy outputs. This feature is very useful, especially during the development and deployment stages.

\begin{figure}
	\centering
 		\includegraphics[width=1.0\linewidth]{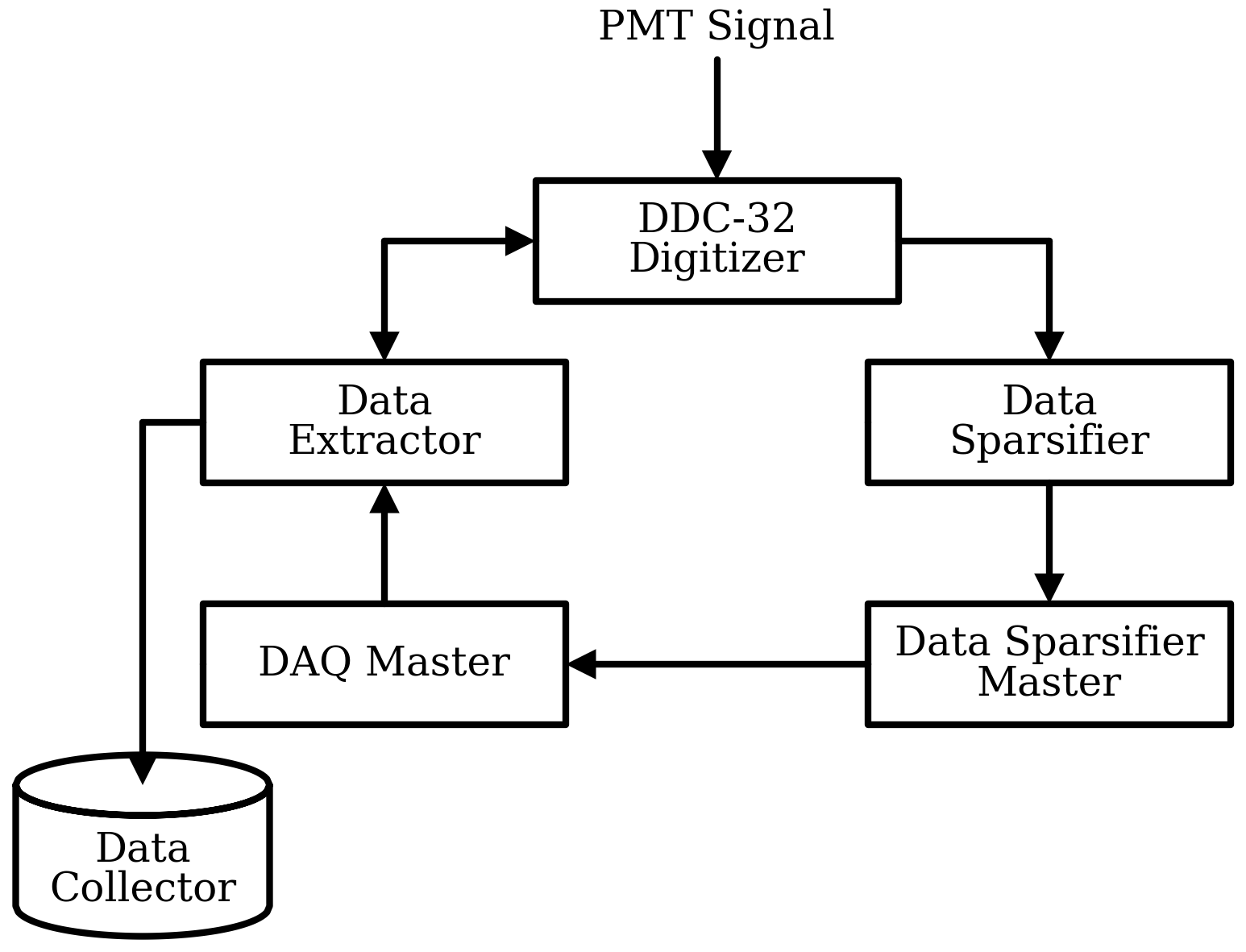}
		\caption{Schematic of the digital signal flow for one PMT channel.}
		\label{fig:daqHardwareOneChannel}
\end{figure}

A schematic for one channel of the digital electronics is shown in Fig.~\ref{fig:daqHardwareOneChannel}.  The DDC-32 digitizes the PMT signal, stores zero-suppressed data in memory, and carries out a basic waveform analysis, discussed in detail in Sec.~\ref{sec:firmware}.  The results of this analysis is sent to one of the Data Sparsifiers.  The information aggregated by the Data Sparsifiers is sent to the Data Sparsifier Master where the decision is made if an event of interest has been observed.  If an event of interest has been observed, the Data Sparsifier Master informs the DAQ Master.  The DAQ Master determines the time window of the event and instructs the Data Extractor to extract the relevant data from the DDC-32s.  The extracted data are stored on one of the Data Collectors.  The Data Extractor, Data Sparsifier, the Data Sparsifier Master, and the DAQ Master use the same hardware (the logic board shown in Fig.~\ref{fig:daqHardware}) with different firmware.

\section{FADR}\label{sec:DAQ}

The design of FADR was based on our experience with LUX \cite{akerib:2011ix,Akerib_2016} and the required calibration rates and data volume \cite{Akerib:2019fml}. The TPC internal source calibration rates are limited by the maximum drift time of 700~$\mu$s. A 150~Hz calibration rate in the TPC results in a 10\% probability of detecting a second calibration event within the drift time of the previous calibration event. Neutron calibrations are carried out to define the nuclear-recoil band in the TPC and to calibrate the response of the OD to neutrons. These calibrations use external neutron sources and a DD neutron generator. External radioactive sources, inserted into the source tubes around the central cryostat, are used to calibrate the TPC, Skin, and OD PMTs. Monthly LED calibrations are performed to monitor the SPHE response of the PMTs. These calibrations can be carried out with rates as high as 4~kHz.  Typical rates used during LED calibrations are 1~kHz for the TPC and Skin PMTs and 0.7~kHz for the OD PMTs.  With these rates, an LED calibration can be carried out in less than 10 minutes.

\begin{figure*}
\centering
\includegraphics[width=1.0\linewidth]{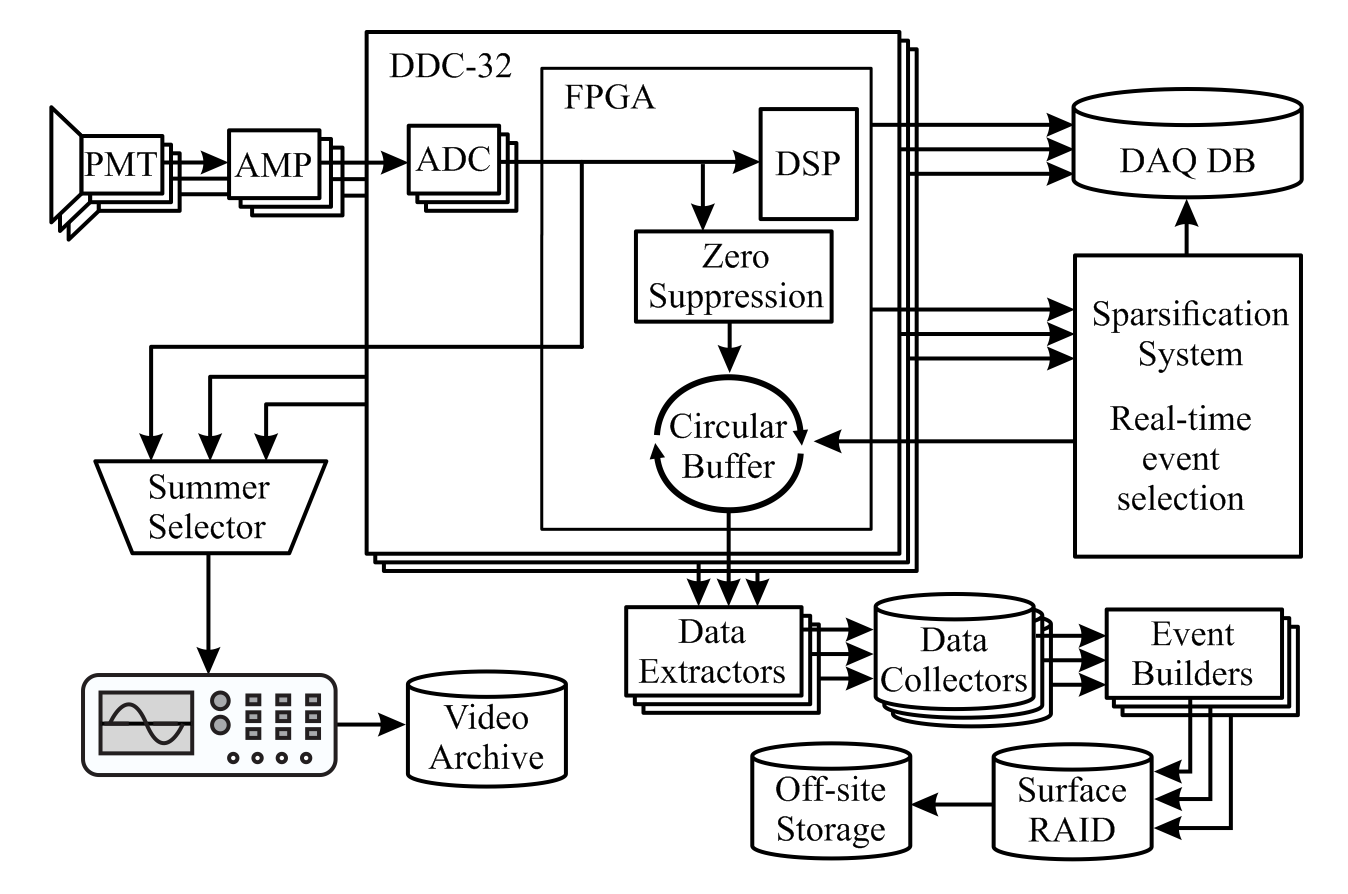}
\caption{Schematic view of the operation of FADR.  Digitized data from the ADCs are stored in circular buffers on the FPGA.  The digitized data are also processed by digital filters (DSP) that provide information to the Sparsification System for real-time event selection.  Once an event of interest has been identified, the waveform samples within the event window are extracted from the circular buffers and used to build events.  Individual or summed waveforms can be viewed in real time on the DAQScope whose video stream is archived.  The results of various monitoring utilities on the FPGA are stored in the DAQ database (DB) and are available to monitor the performance of the detector and the electronics system.}
\label{fig:DAQSchematic}
\end{figure*}

The top-level architecture of FADR system is shown schematically in Fig.~\ref{fig:DAQSchematic}. The DDC-32s continuously digitize the incoming PMT signals and store waveform data in circular buffers. When an event is detected, the Data Extractors collect the data of interest from all DDC-32s and send them to Data Collectors for temporary storage. Event Builders take the data, organized by channels, and assemble the buffers into full event structures for online and offline analysis. The system runs synchronously with one global clock, which is discussed in more detail in Sec.~\ref{subec:TimeStampting}.

\begin{figure*}
\centering
\includegraphics[width=1.0\textwidth]{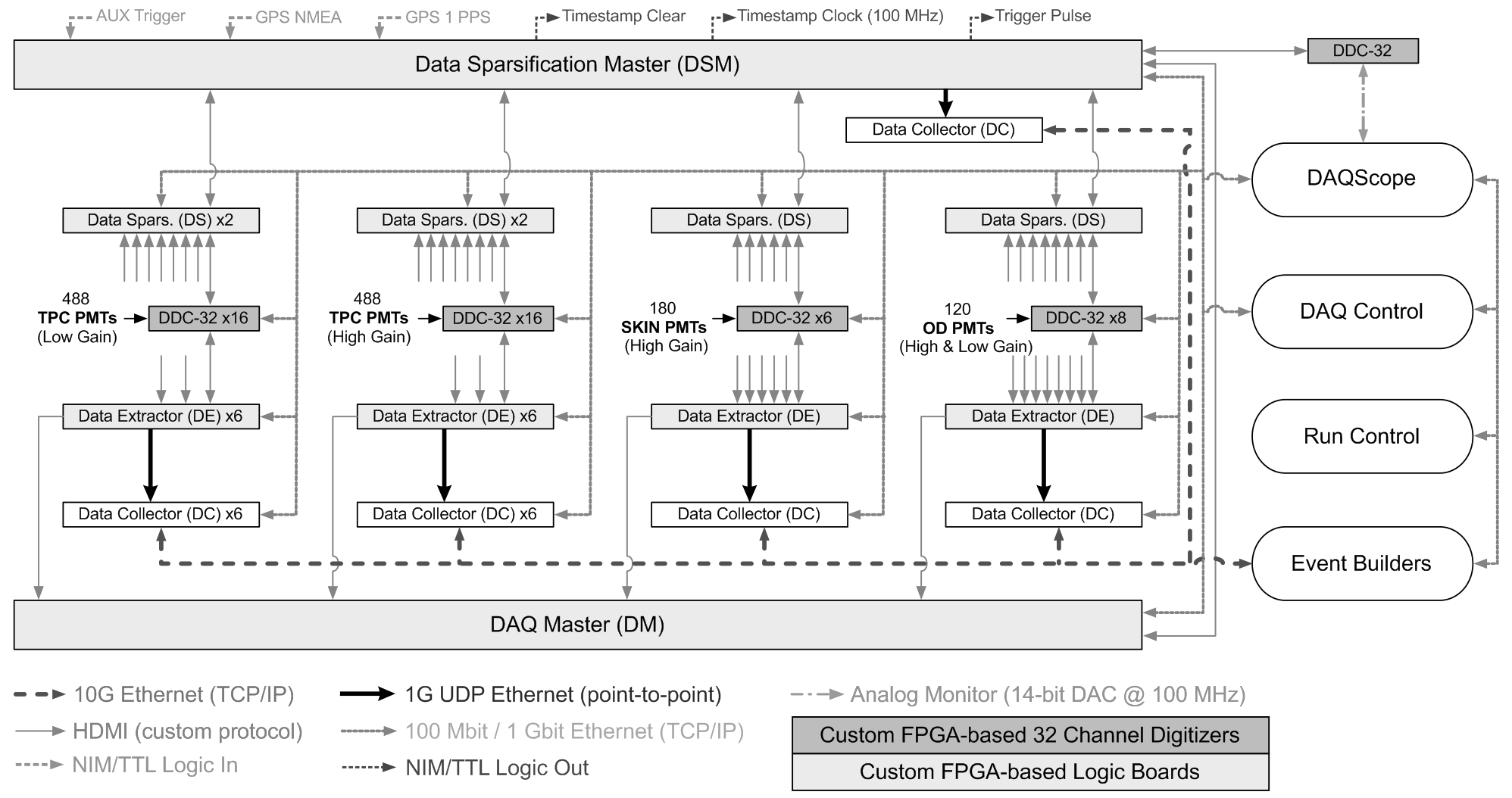}
\caption{Diagram of the architecture of FADR. Groups of DDC-32s capture the amplified and shaped signals from the TPC, Skin, and OD PMTs. The waveforms of interest are extracted from the DDC-32s by the Data Extractors before they are passed to Data Collectors for temporary storage. One additional Data Collector captures reduced sparsification quantities to be merged by the Event Builder with the waveform data in full event files. The DAQ Master coordinates the high-speed operation of FADR when the Data Sparsification Master signals the detection of an event to be preserved. The global clock is distributed over HDMI links.}
\label{fig:FADRArchitecture}
\end{figure*}

A detailed view of FADR is shown in Fig.~\ref{fig:FADRArchitecture}.  The system contains four parallel processing chains, one for the high-gain TPC PMTs, one for the low-gain TPC PMTs, one for the high-gain Skin PMTs, and one for the high- and low-gain OD PMTs.  For the TPC PMTs, up to three DDC-32s are connected to one Data Extractor; the 32 TPC DDC-32s require a total of 12 Data Extractors (see Table~\ref{tbl:HardwareDistribution}).  The five DDC-32s serving the Skin PMTs are connected to one Data Extractor, and the eight DDC-32s serving the OD PMTs are also connected to a single Data Extractor.  The number of DDC-32s connected to a Data Extractor is determined by the expected data volume to be extracted from the DDC-32s and the data transfer limitations between each Data Extractor and its Data Collector.  Details on the distribution of hardware elements across the four parallel processing chains are provided in Table~\ref{tbl:HardwareDistribution}.

\begin{table}
\caption{Hardware distribution across the four parallel processing chains.}
\label{tbl:HardwareDistribution}
\centering
\vspace{0.25cm}
\small
\begin{tabular}{ >{\centering\arraybackslash} p{0.26\linewidth} >{\centering\arraybackslash} 
p{0.14\linewidth} >{\centering\arraybackslash} 
p{0.12\linewidth} >{\centering\arraybackslash} p{0.12\linewidth} >{\centering\arraybackslash} p{0.12\linewidth} }
    \hline
    Processing chain & Channels & DDC-32s & Data Extractors & Data Sparsifiers \\
    \hline
    TPC High Gain & 494 & 16 & 6 & 2 \\
    TPC Low Gain & 494 & 16 & 6 & 2 \\
    Skin & 131 & 5 & 1 & 1 \\
    OD & 240 & 8 & 1 & 1 \\
    Sensors & 21 & 1 & 1 & 1 \\
    \hline
\end{tabular}
\end{table}

\begin{figure}
\centering
\includegraphics[width=0.9\linewidth]{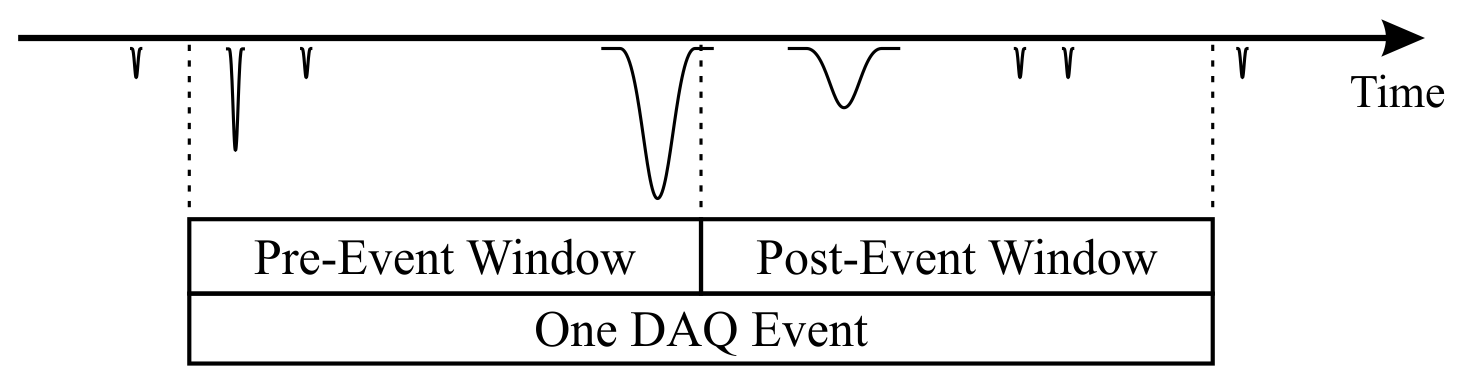}
\caption{Event definition. The time range of an event is defined by two parameters: the pre-event and the post-event window. These in general have different widths and differ for different acquisition modes. The vertical line in the center of the figure is the time at which the Data Sparsifier Master detects an event of interest.  }
\label{fig:eventDefinition}
\end{figure}

The event selection is made by the data sparsification system.  Data Sparsifiers collect information about the waveform analysis that is carried out on each DDC-32, described in detail in Sec.~\ref{subsec:DSP}, and send it to the Data Sparsification Master. The Data Sparsifier Master makes the final event selection, based on the trigger information provided by the four trigger chains shown in Fig.~\ref{fig:FADRArchitecture} and the external triggers received by the Data Sparsifier Master.  When the Data Sparsifier Master detects an event of interest, it contacts the DAQ Master which instructs the Data Extractors to extract the relevant data from the DDC-32s.  The time range of waveform data extracted from the DDC-32s is defined by the pre- and post-event windows, shown schematically in Fig.~\ref{fig:eventDefinition}.  

The Data Extractors, Data Sparsifiers, DAQ Master, and Data Sparsifier Master use the same hardware, shown in Fig.~\ref{fig:daqHardware}(right), but different firmware.  The boards are connected via a total of 114 HDMI cables. The HDMI cables are used as the physical connection, carrying custom-developed communication protocols. The physical HDMI cable has five truly differential pairs (twisted and shielded) and four single-ended wires which are configured as two lower-performance differential pairs. All seven pairs are connected directly to the FPGA's LVDS serial/deserializer (IOSERDES) elements. One true differential pair is dedicated to the precise distribution of the global 100-MHz clock to all the boards. The two lower-performance pairs are used as an 800~Mbps (2$\times$400~Mbps) control channel, while the remaining four true pairs are used as 3.2~Gbps (4$\times$800~Mbps) data sparsification and data transfer channels.  Details on error checking of the data is provided in Sec.~\ref{subsec:dataOffloading}.

\subsection{Digitization}\label{subec:Digitization}

\begin{figure*}
\centering
\includegraphics[width=1.0\linewidth]{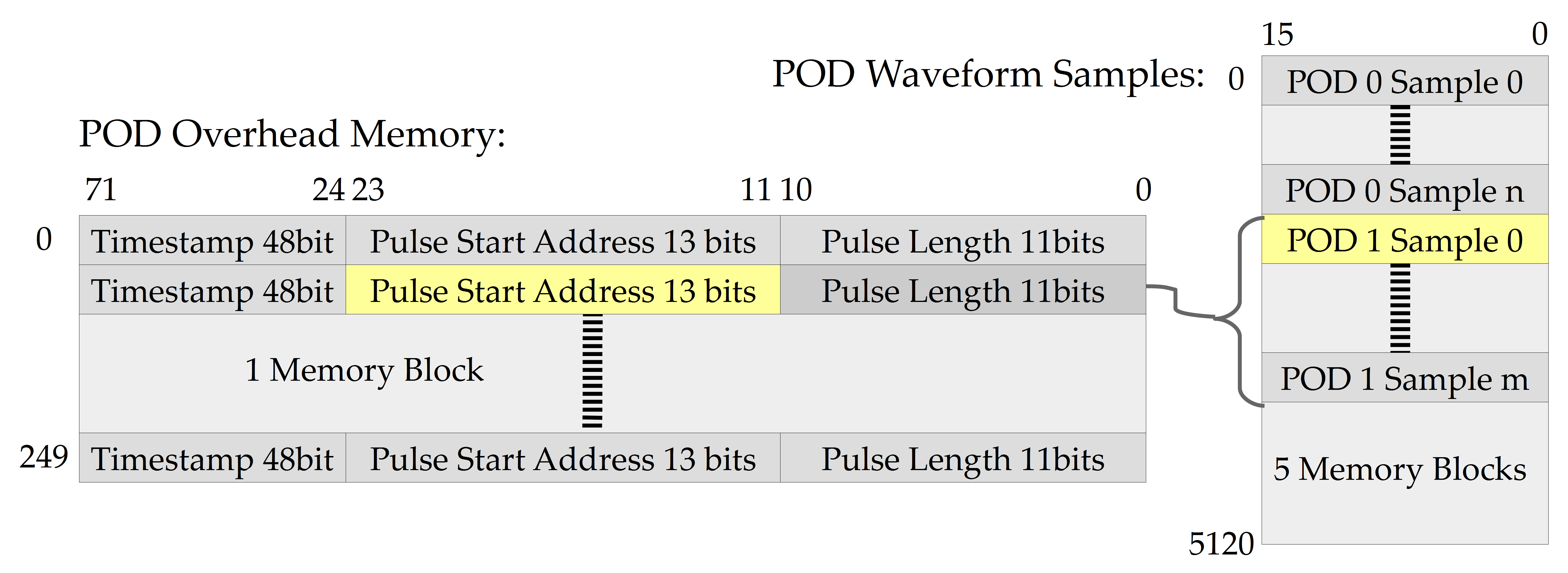}
\caption{The memory organization of the POD waveform storage on the FPGA for a single buffer (out of two) of a single channel. The separation of POD header and POD samples improves the performance and ease of extracting information from memory.}
\label{fig:DAQMemory}
\end{figure*}

The DDC-32s sample at 100 MHz with 14-bit resolution over a 2-V range. A 0.8-V offset is applied to the input signal to provide an effective dynamic range between -1.8~V and +0.2~V. During normal operation, the DDC-32s collect waveforms in a Pulse Only Digitization (POD) mode, discussed in more detail in Sec.~\ref{subsec:zeroSuppressionFirmwware}, which reduces the raw waveform data volume by a factor of up to 50~\cite{akerib:2011ix}. The POD waveforms are stored in dual-buffered memory that is divided into sections that separately hold the header information and the actual POD samples, as shown in Fig.~\ref{fig:DAQMemory}. Separate POD header and payload memories improve the performance of extracting waveform data when the Data Sparsifier Master detects an event of interest~\cite{Druszkiewicz:2015}.

\subsection{Data Extraction}\label{subec:DataExtraction}

Each Data Extractor can serve up to 14 DDC-32s using LVDS links.  The number of DDC-32s connected to each Data Extractor and Data Sparsifier is based on the maximum expected data volume.  For the TPC PMTs, the S2 signals produce a significantly higher data volume per event compared to the Skin and OD PMTs which only see S1-like signals.  Each Data Extractor connects to one Data Collector over a dedicated Gigabit Ethernet connection

\begin{table}
\caption{Key parameters of the Data Collectors.}
\label{tbl:DC}
\centering
\vspace{0.25cm}
\small
\begin{tabular}{ p{0.25\linewidth} p{0.65\linewidth}} 
\hline
{\bfseries Processor:} & Intel Xeon E3-1270V3 3.5GHz Quad-Core \\
{\bfseries Motherboard:} & ASUS P9D-V ATX \\
{\bfseries Memory:}	& 16GB Kingston DDR3 SDRAM ECC \\
{\bfseries NIC:} & Intel Ethernet Server Adapter I350-T2 \\
{\bfseries Disk 1:} & SAMSUNG 860 Pro Series 1TB SSD \\
{\bfseries Disk 2:} & Western Digital RE 4TB 7200 RPM \\
{\bfseries Case:} & NORCO RPC-270 2U Server Case \\
{\bfseries Hot Swap:} & ICY DOCK 3.5" and 2.5" SATAIII 6Gps HDD Rack Tray \\
\hline
\end{tabular}
\end{table}

\begin{table}
\caption{Summary of the performances of the DAQ links and their measured utilization levels during WIMP search.}
\label{tbl:DAQLinks}
\centering
\vspace{0.25cm}
\small
\begin{tabular}{ p{0.25\columnwidth} >{\centering\arraybackslash} p{0.30\columnwidth} >{\centering\arraybackslash} p{0.30\columnwidth} } 
    \hline
{\bfseries Link} & {\bfseries Sustained Maximum Performance} & {\bfseries Measured Maximum Usage} \\
\hline
LVDS over HDMI & 190 MB/s & 0.9 MB/s \\
Gigabit UDP & 115 MB/s & 4.3 MB/s \\
\hline
\end{tabular}
\end{table}

The Gigabit Ethernet links between the Data Extractors and the Data Collectors utilizes the User Datagram Protocol (UDP). UDP is used since it was significantly easier to implement a pure VHDL 1G UDP stack compared to a 1G TCP/IP stack.  To guarantee that we only record uncorrupted data, we implemented custom checksumming at the event and packet level. Dedicated ethernet links in a point-to-point topology are used to eliminate congestion and packet loss. All UDP packets formed by each Data Extractors FPGA are sent over a single Gigabit Ethernet cable to a dedicated network port on the corresponding Data Collector.

\subsection{Data Collectors}\label{subec:DataColletor}

The Data Collectors are rack mountable 2U servers with the parameters shown in Table~\ref{tbl:DC}.  After fine tuning the network interface card driver parameters, we were able to transfer data to each Data Collector at a rate of 109.8~MB/s without any data loss or corruption. Table~\ref{tbl:DAQLinks} summarizes link performances and their utilization.

\subsection{Event Builders}\label{subec:EventBuilders}

The data stored on the Data Collectors are processed by Event Builders. Five Dell PowerEdge R530/R540 servers are used for event building.  The data flows from the Data Collectors to the Event Builders via a dedicated 10 Gb/s switched network.

For a given acquisition, every newly acquired Data Collector file will contain the same number of events, typically between 50 and 250. The Event Builders build events by polling the SSDs on all Data Collectors and collecting all information associated with a given event. Each event file contains the same number of events as the individual Data Collector files.  Each individual Event Builder file is called a sequence file and during a typical run, 500 to 2,000 sequence files are collected.  Each Event Builder has 72 TB of disk space where the sequence files are stored, before being transferred to two RAID arrays, located on the surface at SURF. From there, the data is transferred to the US data center at the National Energy Research Scientific Computing Center (NERSC) \cite{NERSC} at Lawrence Berkeley National Laboratory.  The RAID arrays at SURF have sufficient storage capacity for a full month of data taking so that data collection at SURF can continue even during periods when data transfer to NERSC is not possible.  The data can also be transferred to the data center in the United Kingdom \cite{LzUK}. This option is important if NERSC will not be available for extended periods of time.

\subsection{Time Stamping}\label{subec:TimeStampting}

FADR runs off one global 100-MHz clock source, distributed via one of the differential pairs of the HDMI cables. The 48-bit time-stamping counters are implemented using dedicated high-performance DSP48 slices on the FPGA \cite{Xilinx:DSP48}.  These counters are reset at the beginning of each run. The 48-bit counters, running at a 100-MHz increment rate, overflow after 32 days, long after the typical acquisition length of between 2 and 4 hours.

\begin{figure*}
\centering
\includegraphics[width=0.90\linewidth]{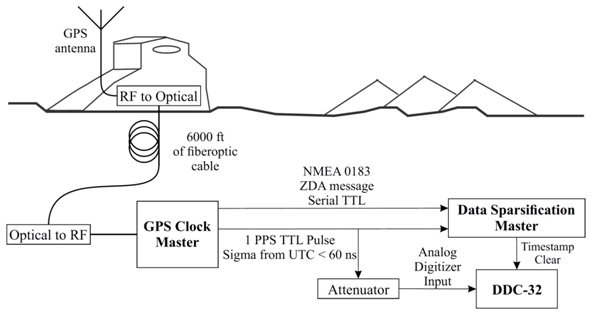}
\caption{Schematic of the GPS timing system.}
\label{fig:gpsTiming}
\end{figure*}

Time derived from GPS is integrated into the data stream.  A schematic of the GPS timing system is shown in Fig.~\ref{fig:gpsTiming}.  A GPS antenna is mounted on the roof of the administration building at SURF.  Optical Zonu GPS-over-Fiber extenders are used to get the GPS signal to the underground master clock.  A Trimble Mini-T GG Multi-GNSS Disciplined Clock is used as our GPS master clock. It is designed to provide the 1 pulse-per-second (PPS) signal with a 15~ns (1-sigma) accuracy. When an acquisition is started, the Data Sparsifier Master resets FADR's internal 48-bit timestamping counter to zero and aligns it with the rising edge of the PPS pulse coming from the GPS master clock. The PPS pulse train is digitized, using a spare ADC channel, along with the PMT signals.

\subsection{FADR Control}\label{subec:DaqControl}

The configuration, acquisition control, and monitoring is done using the DAQ Control computer. For improved robustness, the DAQ Control computer communicates with all elements of FADR on an isolated local-area network, using Remote Python Calls (RPyC) as the communication framework~\cite{RPyC}.

Run Control (RC) is the web-based user interface to FADR.  The user can start and stop acquisitions, schedule new acquisitions, and monitor event building and data flow.  For each acquisition, the user can select pre-defined configuration settings.  Only one user can be operator, but multiple users can view the Run Control interface at the same time.  Run Control uses Internet Communication Engine (ICE) middleware as the communication framework~\cite{Henning:2004}.

\section{Firmware}\label{sec:firmware}

\subsection{Zero suppression}\label{subsec:zeroSuppressionFirmwware}

\begin{figure}
\centering
\includegraphics[width=0.75\linewidth]{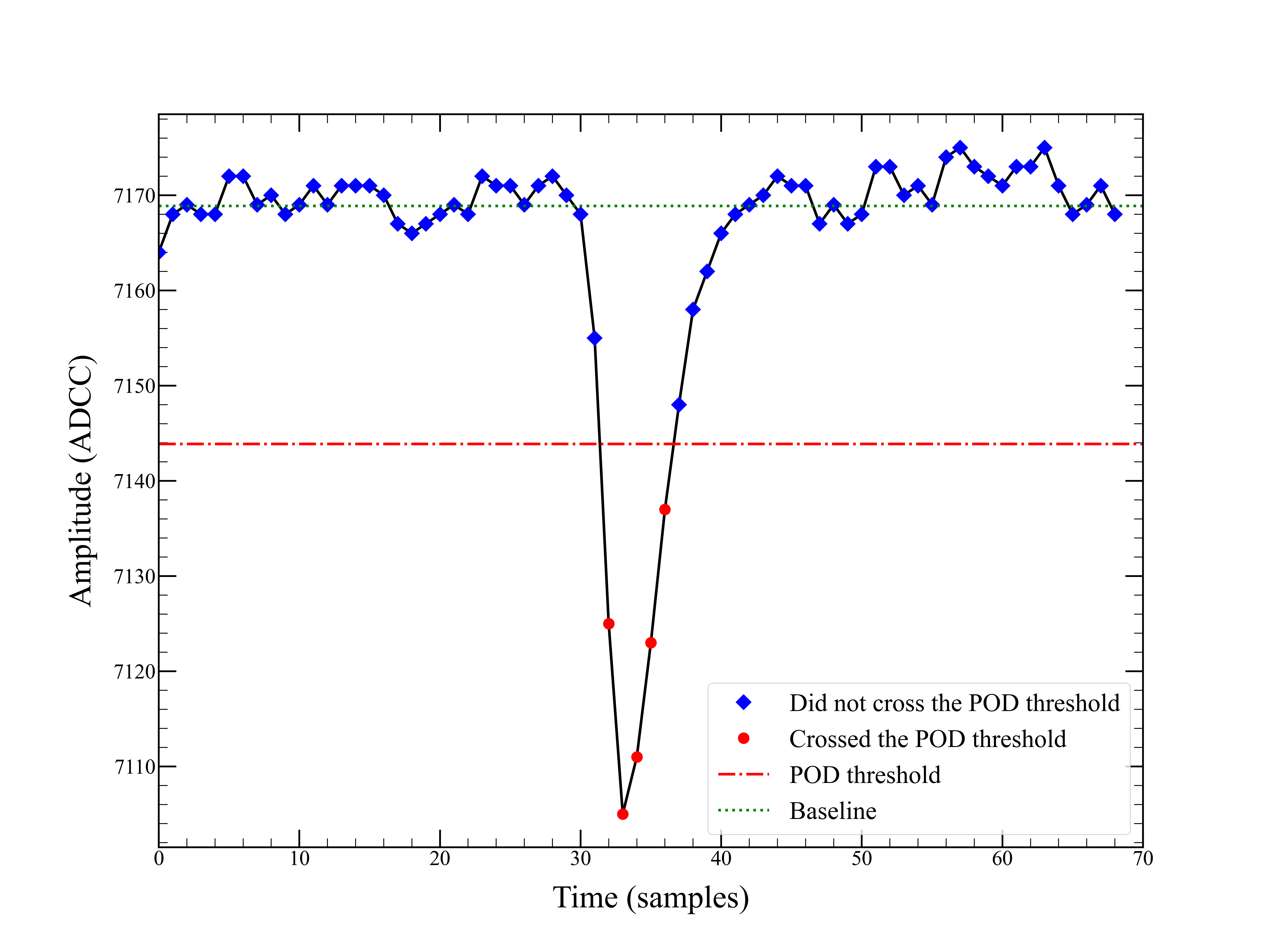}
\caption{Example of a POD, including the pre- and post-samples.  This POD was generated with a POD threshold of 25 ADCC.}
\label{fig:OnePod}
\end{figure}

During normal operation, the baseline of the waveform is defined as the rolling average of the 32 ADC samples before the current sample. A deviation of a sample from the baseline by more than the user-defined threshold, which is constant during a run, initiates the process that stores the waveform in memory. This process is called Pulse Only Digitization and the stored waveform is called a POD.  In addition to the region of the waveform above the POD threshold, the 32 samples before the POD threshold crossing (the pre-samples) are stored as well as the 32 samples after the waveform returns to values below the POD threshold (the post-samples).  If another threshold crossing occurs during the post-sample period, the overlapping PODs are merged.  An example of a POD is shown in Fig.~\ref{fig:OnePod}.

\subsection{Data Sparsification}\label{subsec:DSP}

\begin{figure*}
\centering
\includegraphics[width=0.75\linewidth]{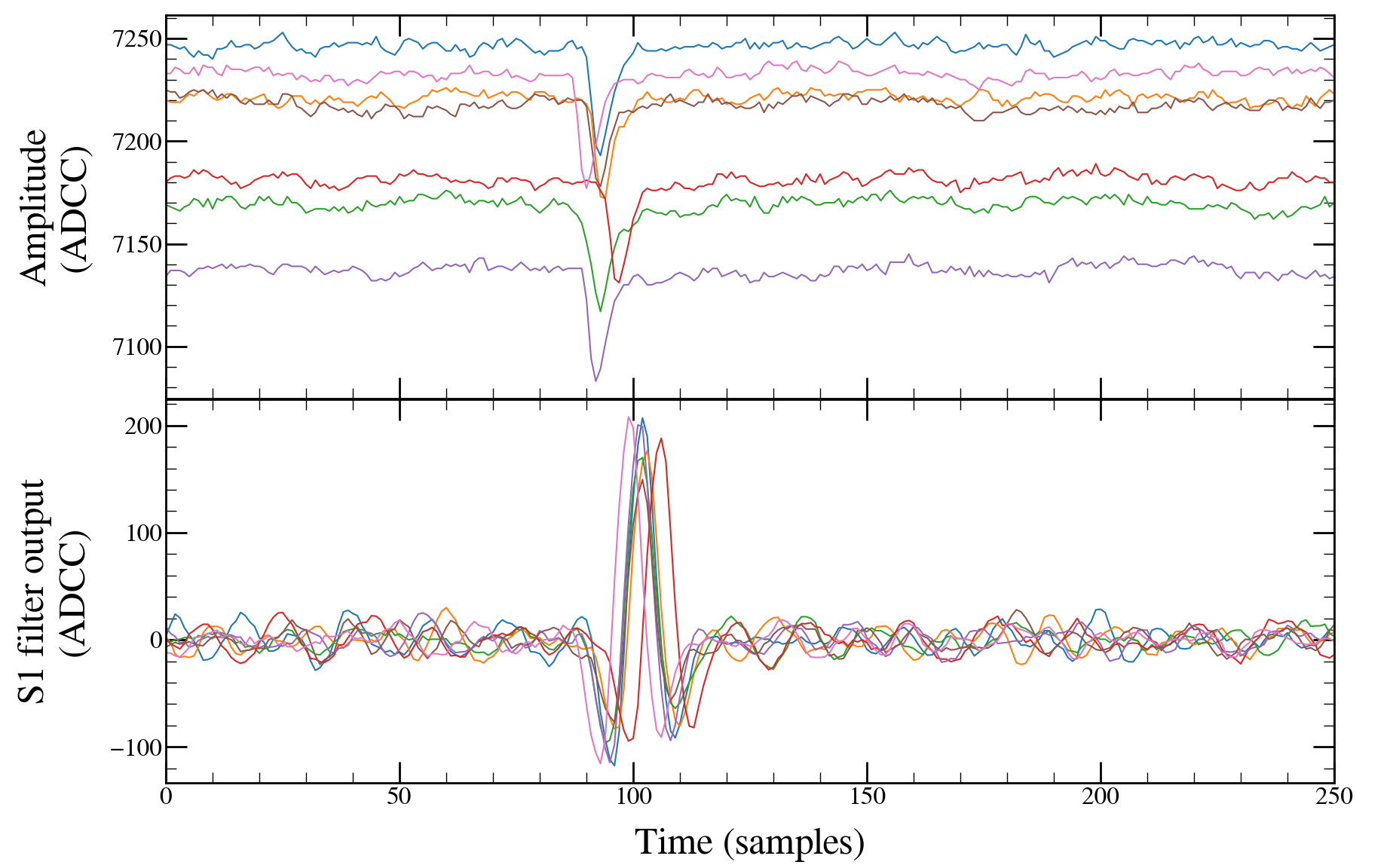}
\caption{Top: SPHE calibration waveforms from seven PMTs.  Bottom: The S1 filter output of the seven waveforms shown in the top part of this figure are processed by an S1 filter with a central lobe of 4 samples.}
\label{fig:beforeAndAfterS1Filter}
\end{figure*}

To achieve a high SPHE efficiency, the effect of slow baseline variations must be reduced.  In addition, the baseline of different channels will be slightly different due to small DC offsets in the signal chain. This effect is illustrated in Fig.~\ref{fig:beforeAndAfterS1Filter}(top) which shows SPHE waveforms for seven PMTs, obtained with LEDs.  The baseline variations of the different PMTs are clearly visible. In order to reduce the sensitivity to baseline variations, the waveforms are processed by digital Finite-Impulse-Response (FIR) filters \cite{Hogenauer:1981} which perform signal integration and baseline subtraction.

\begin{figure}
\centering
\includegraphics[width=0.75\linewidth]{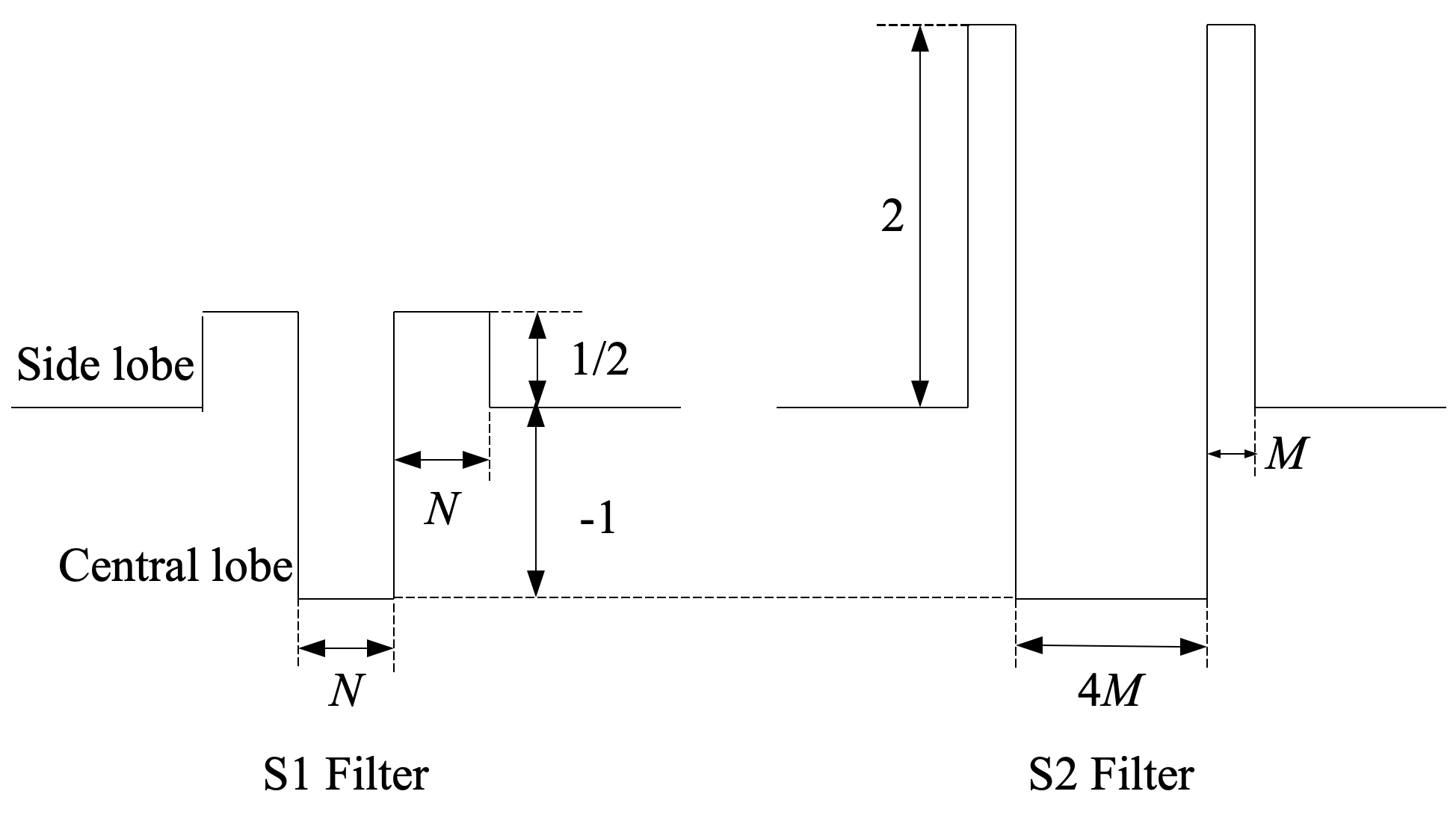}
\caption{The FIR filters used to process the incoming waveforms.  The S1 and S2 filters are used to look for S1 and S2 signals, respectively.}
\label{fig:S1S2Filters}
\end{figure}

The FIR filters used in FADR are called the S1 and S2 filters and are shown schematically in Fig.~\ref{fig:S1S2Filters}.  For each filter, the central lobe is chosen to be wider than the widths of the pulses of interest in order for the filter output to be proportional to the pulse area. The FIR filters process the raw waveforms, not the PODs discussed in Section~
\ref{subsec:zeroSuppressionFirmwware}.

The filter output $F$ for the S1 filter at time $t$ is calculated from the previous 3$N$ samples of the waveform $a$:

\begin{equation}
    F=\frac{1}{2}\sum\limits_{i=-3N}^{-2N-1}{a\left( i \right)}-\sum\limits_{i=-2N}^{-N-1}{a\left( i \right)}+\frac{1}{2}\sum\limits_{i=-N}^{-1}{a\left( i \right)}
    \label{eq:S1Filter}
\end{equation}

\noindent The total S1 filter width is three times the width of the central lobe. The effectiveness of the S1 filter is demonstrated in Fig.~\ref{fig:beforeAndAfterS1Filter}(bottom) which shows the S1 filter output for the waveforms shown in the Fig.~\ref{fig:beforeAndAfterS1Filter}(top). The baseline of the filter outputs are centered around 0 and the amplitudes are proportional to the area of the SPHE waveforms above their baseline.

The wider S2 filter uses significantly more LookUp-Table (LUT) resources on the FPGA. To reduce the LUT resources used by S2 filters, the side lobes are four times narrower than the central lobes.  The filter value $F$ for the S2 filter at time $t$ is calculated from the previous 6$M$ samples of the waveform $a$:

\begin{equation}
    F=2\sum\limits_{i=-6M}^{-5M-1}{a\left(i \right)}-\sum\limits_{i=-5M}^{-M-1}{a\left( i \right)}+2\sum\limits_{i=-M}^{-1}{a\left( i \right)}
    \label{eq:S2Filter}
\end{equation}

\noindent The total S2 filter width is 1.5 times the width of the central lobe.

The S1 filters operate with central lobes of up to 16 samples (160~ns integration time) while the S2 filters operate with central lobes of up to 512 samples (5.12~$\mu$s integration time).  The filter parameters used for each of the four trigger systems shown in Fig.~\ref{fig:FADRArchitecture} can be different.

Event selection is based on the number of S1 or S2 filter threshold crossings within a configurable coincidence window.  The S1 and S2 filters are continuously applied to the waveforms and information about which PMTs are above the filter threshold is communicated to the Data Sparsifiers.  Each time a PMT has a filter output that is above threshold, it will contribute to the total multiplicity for the next $C$ samples, where $C$ is the width of the coincidence window in samples. The multiplicity information is shared with the Data Sparsifier Master where the final event selection is made. The decision to preserve the current event is sent to the Data Acquisition Master. The time range of the waveform data extracted from the DDC-32s is defined by the pre-event window and the post-event window, shown schematically in Fig.~\ref{fig:eventDefinition}.  The lengths of the pre- and post-window are configurable parameters that can be adjusted for each acquisition.  During WIMP Search acquisitions, the pre-event window was 2~ms and the post-event window was 2.5~ms.

\begin{figure*}
\centering
\includegraphics[width=1.0\linewidth]{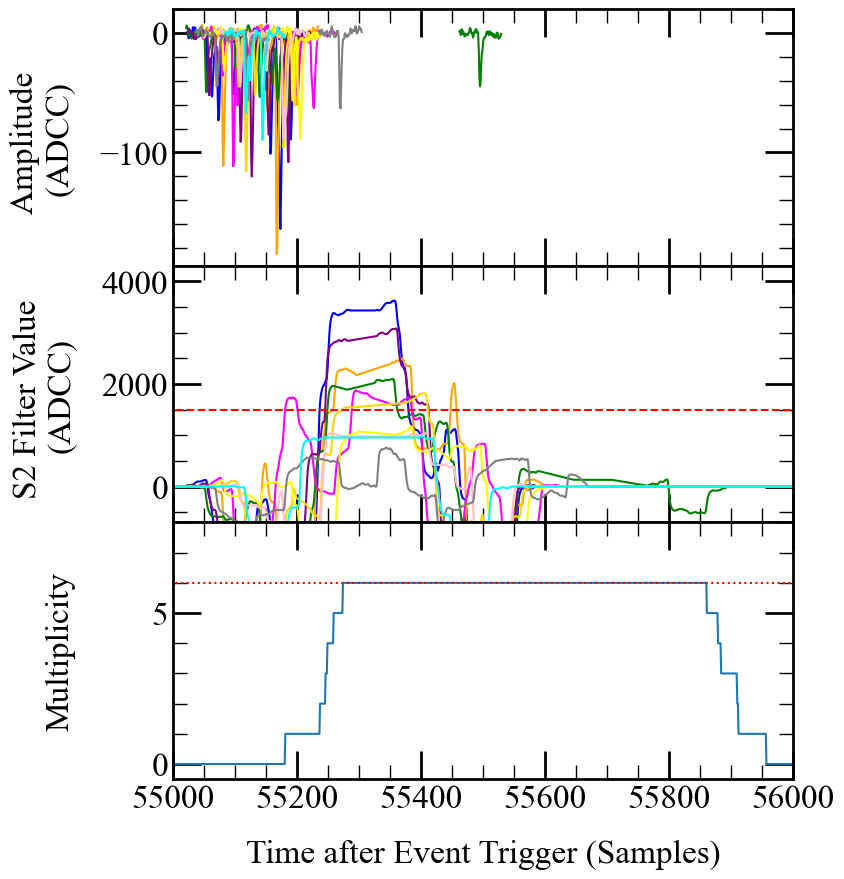}
\caption{An example of a S2 pulse with 190 photons detected across 50 PMTs.  The top figure shows the individual baseline subtracted waveforms of these 50  PMTs.  The middle figure shows the S2 filter values for the ten brightest PMTs of this event.  The red dashed line indicates the S2 filter threshold.  Six PMTs cross the S2 filter threshold. The bottom figure shows the multiplicity array that is used to make the event selection. The red dashed line indicates the multiplicity condition for the S2 trigger.  At sample 55,274, the multiplicity condition is met.}
\label{fig:ExampleS2}
\end{figure*}

After an event is captured, a post-event holdoff is applied.  During the holdoff period no triggers initiate event capture.  The post-event holdoff is applied to reduce the probability of retriggering on the tails of a large S2 pulse.  The holdoff period used during LZ's first Science Run (SR1) was 2 ms.  
The event selection parameters used during SR1 are listed in Table~\ref{tbl:triggerParameters}.  Figure~\ref{fig:ExampleS2} shows an example of the S2 trigger operation.

\begin{table}
\caption{Event selection parameters used during SR1.  Shown are filter type, width of the central lobe, filter threshold, width of the coincidence window, and the required multiplicity.}
\label{tbl:triggerParameters}
\centering
\vspace{0.25cm}
\small
\begin{tabular}{ >{\centering\arraybackslash} p{0.1\linewidth} >{\centering\arraybackslash} p{0.17\linewidth} >{\centering\arraybackslash} p{0.17\linewidth} >{\centering\arraybackslash} p{0.17\linewidth} >{\centering\arraybackslash} p{0.17\linewidth} }
    \hline
    System/ Filter & Filter Width (samples) & Threshold (ADCC) & Coincidence Width (samples) & Multiplicity \\ 
    \hline
    TPC/S2 & 244 & 1500 & 500 & 6 \\
    OD/S1 & 15 & 3000 & 32 & 15 \\
    \hline
\end{tabular}
\end{table}

Much of the information on which a given event selection was made, such as the PMT channels that contributed to the event selection, is stored alongside the waveform data in the event files. This allows for cross-checking and verification of the performance of the system offline.  When an event is selected for readout, the waveforms from all PMT channels are included in the event, independently of which detector volume generated the trigger.  

The two major waveform-selection modes used are summarized in Table~\ref{tbl:triggerModes}.  The signals in the Skin and the OD are S1-like signals. The Skin and OD use the S1 trigger mode to trigger on these detector volumes. In order to be sensitive to S2-only events in the TPC, the TPC trigger uses the S2 trigger mode. 

\begin{table}
\caption{Summary of two major waveform-selection modes.}
\label{tbl:triggerModes}
\centering
\vspace{0.25cm}
\small
\begin{tabular}{  p{0.25\linewidth}  p{0.65\linewidth} } 
    \hline
    Trigger Mode & Description \\
    \hline
    S1 mode & Detection of coincident S1-like signals across selected channels. No fiducialization. \\
    S2 mode & Detection of coincident S2-like signals across selected channels. Fiducialization in the $x$,$y$ plane is possible for the TPC PMTs. \\
    \hline
\end{tabular}
\end{table}

\subsection{Rate and Noise Monitoring}\label{subsec:noiseMonitor}

The FPGA on each DDC-32 continuously monitors various rates that are important to ensure the proper operation of the detector and the electronics.  These rates are sensitive to electronic noise, sensitive to light emission in the detector, and/or sensitive to the various trigger rates.  The extracted rates are written to a MySQL database that can be accessed using tools such as Grafana~\cite{grafana} and Tableau~\cite{tableau} to look at long-term trends in these rates and set alarms on short-term changes.  The following rates are captured every 10 s:
\begin{itemize}
    \item S1 filter threshold crossing rates for each PMT channel.  Two different S1 filter parameters are used for each PMT channel: one set of parameters are chosen to make the filter rates sensitive to electronics noise, the other set of filter parameters is selected to monitor the SPHE rates in the PMTs.  These S1 filters operate in parallel to the S1 filters that are used for event selection.
    \item S2 filter threshold crossing rates for each PMT channel.  The parameters of this filter can be different from the parameters of the S2 filters that are used for event selection.
    \item POD threshold crossing rates for each PMT channel.
    \item The number of samples above the POD threshold per second for each PMT channel.
    \item The individual trigger rates for all detector systems, including the external triggers.
\end{itemize}

\begin{figure}
\centering
\includegraphics[width=\linewidth]{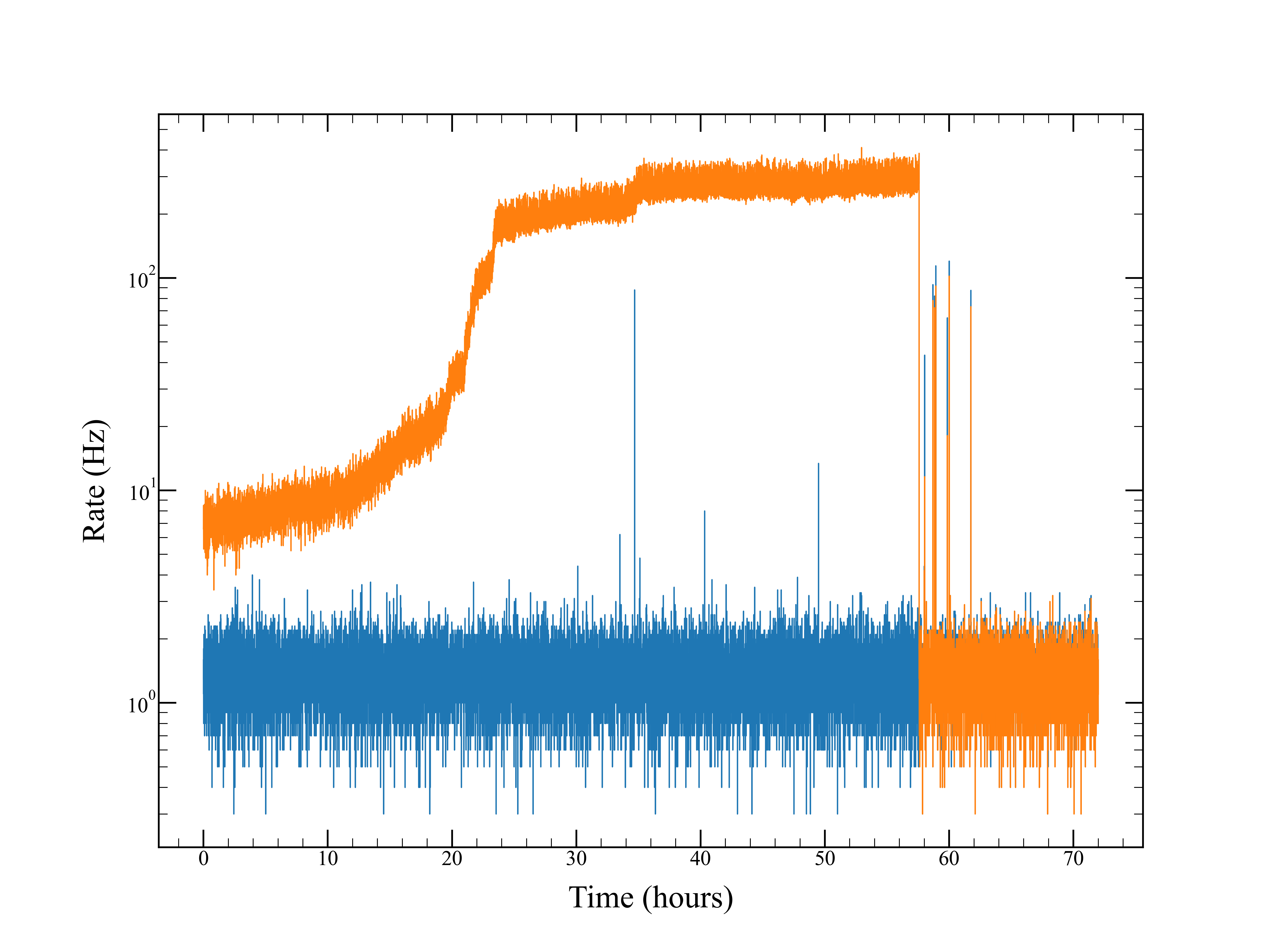}
\caption{Noise (top trace) and SPHE (bottom trace) rates used to monitor the performance of the detector, the electronics, and FADR.  The rates shown are the rates for one TPC PMT during a three-day period.}
\label{fig:DAQDBRates}
\end{figure}

An example of the use of the two S1 monitoring rates for a TPC PMT is shown in Fig.~\ref{fig:DAQDBRates}.  During the period shown in Fig.~\ref{fig:DAQDBRates}, a grounding problem developed with one of the amplifiers.  This caused the noise rate in the channels associated with this amplifier to increase significantly (top trace). The impact of the added noise was sufficiently small that it did not impact the measured SPHE rates (bottom trace).  At 58 hours, the connection of the grounding braid to the amplifier was retightened and the noise rate returned to normal value.

\section{Event Selection}\label{sec:eventSelection}

During normal operation, the event selection is based on multiple trigger sources.  In this section we will consider the detector and the external triggers separately.

\subsection{WIMP-search Triggers}\label{subsec:WimoSearchTriggers}

During normal WIMP-search operation, the detector-based event selection is the logical OR of the following triggers:

\begin{itemize}
    \item The TPC S2 trigger.  This trigger is based on an analysis of the S2 filter response of the high-gain signals of the top TPC PMTs.
    \item The Skin S1 trigger.  This trigger is based on an analysis of the S1 filter response of the high-gain signals of the Skin PMTs.
    \item The OD S1 trigger.  This trigger is based on the S1 filter response of the high-gain signals of the OD PMTs.
\end{itemize}

\noindent The Skin and OD triggers are used to monitor the health of these detector volumes during WIMP-search operation.  The trigger settings for the Skin and OD are such that the corresponding trigger rate is a small fraction of the TPC trigger rate.  When the rate of a particular trigger is too high, a downscale factor can be applied to the trigger source before an event selection decision is made.  The length of the pre- and post-event window around the trigger time is independent of the trigger source.  If additional triggers occur during the post-event window, that information will be captured in the data stream, but it will not change the length of the post-event window.

To monitor the background, the health of the detector, and FADR, the following additional external triggers are used:

\begin{itemize}
    \item A random trigger.  This trigger operates with a random rate of around 4 Hz and provides an unbiased sample, allowing us to monitor the background in the detector and the efficiency of the S1 and S2 triggers. 
    \item The GPS trigger.  This trigger runs at exactly 1 Hz, independent of detector signals, to monitor detector health.  This trigger also allows us to measure the live time of FADR online and offline.
\end{itemize}

\subsection{Calibration Triggers}\label{subsec:calibrationTriggers}

External triggers are used primarily during certain calibrations.  The following external triggers are used:

\begin{itemize}
    \item DD trigger. This trigger is generated by the pulsed neutron generator.
    \item LED triggers. Multiple LED systems are used to calibrate and monitor the health of the PMTs \cite{Turner_2021}.  Each of these systems generates a trigger signal when the LEDs are fired. 
\end{itemize}

\noindent The calibration trigger is the logical OR of the DD and LED triggers.

\section{Event Size}\label{sec:eventSize}

The event size is defined as the ratio of the size of the event files and the number of events contained within them.  During WIMP search the event size is about 0.9~MB.  The size of each event is dominated by the width of the S2 signals in the TPC.  During LED calibrations of the TPC PMTs, only S1-like signals are observed, and the measured event size is about 60~kB.  During neutron calibrations, the event size is 2.1~MB due to the multiple scattering of these neutrons in the LXe and the corresponding multiple S2 signals. The measured event size for other calibration modes are listed in Table~\ref{tbl:calRates}.

\begin{table}
\caption{Examples of calibrations modes, count rates, and average event sizes.  For comparison, the typical rate during WIMP search is 15~Hz with an event size of 0.9~MB.}
\label{tbl:calRates}
\centering
\vspace{0.25cm}
\small
\begin{tabular}{ >{\centering\arraybackslash} p{0.2\linewidth} >{\centering\arraybackslash} p{0.2\linewidth} >{\centering\arraybackslash} p{0.2\linewidth} >{\centering\arraybackslash} p{0.2\linewidth} }
    \hline
    Calibration Source & System & Count Rate (Hz) & Event Size (MB) \\ 
    \hline
    $^{83m}$Kr & TPC & $<$150 & 0.8 \\
    $^{22}$Na & TPC/Skin/OD & 10 & 0.9 \\
    $^{57}$Co & Skin & 135 & 0.2 \\
    DD neutrons & TPC  & $<$150 & 2.1 \\
    AmLi neutrons & TPC/OD & 55 & 3.0 \\
    LEDs & TPC/Skin/OD  & 1,000 & 0.06 \\
    \hline
\end{tabular}
\end{table}

\section{Extended Functionality}\label{sec:extraFunctions}

Since we are in full control of the firmware of FADR, we developed additional features that enhance our ability to test, troubleshoot, monitor, and expand the information on which event selection is based.  In this section, a few examples are discussed.

\subsection{Arbitrary Waveform Injection}

\begin{figure*}
    \centering
    \includegraphics[width=0.9\linewidth]{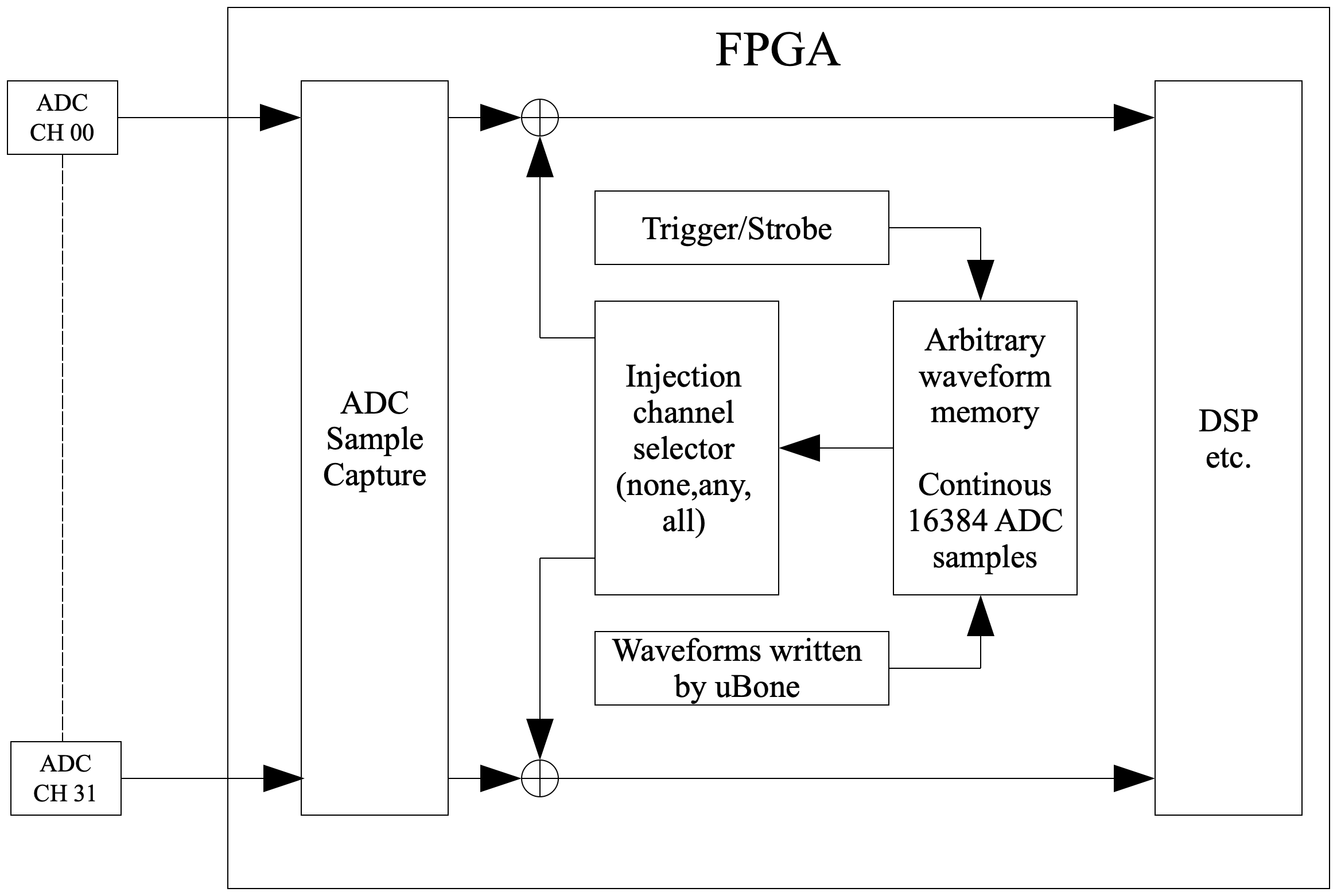}
    \caption{Schematic of the operation of arbitrary waveform injection.}
    \label{fig:arbWaveformInkection}
\end{figure*}

Firmware was developed to allow arbitrary waveform injection at the front-end of the DDC-32.  This mode of operation is shown schematically in Fig.~\ref{fig:arbWaveformInkection}. The injected waveforms are injected after the ADCs and the measured response to the injected waveforms thus includes the real ADC noise. 

We have used this mode of operation in various ways.  For example, by injecting well defined pulses at the front end, we tested the logic of the Data Sparsifier system.  By changing the properties of the injected pulses (e.g. the pulse area and/or the pulse multiplicity) we verified that the Data Sparsifier system was treating different pulses properly.

\subsection{DAQScope}\label{subsec:daqScope}

\begin{figure*}
\centering
\includegraphics[width=1.0\linewidth]{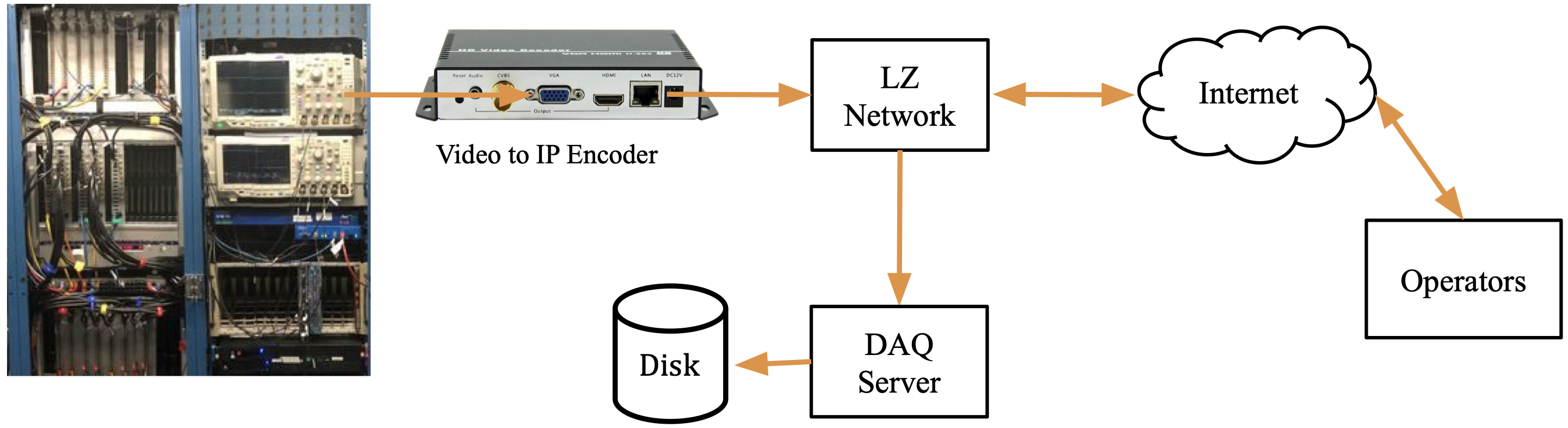}
\caption{Key elements of DAQScope.  Two spy outputs are displayed on an oscilloscope in one of the electronics racks.  The graphics output of the oscilloscope is connected to a Video-to-IP encoder.  This allows the screen of the oscilloscope to be monitored by local and remote users.  An IP-to-Video decoder is used to provide a permanent display of the image on the oscilloscope on a large display in the control room.}
\label{fig:daqScope}
\end{figure*}

DAQScope was developed to allow us to monitor individual signals connected to the DDC-32s without the need to move/unplug/replug any signal cables. The key elements of DAQScope are shown in Fig.~\ref{fig:daqScope}.  The operator can look at any one or a sum of selected PMT channels on a regular oscilloscope, connected to one of the spy outputs of the DDC-32s.  The multiplexing/summing and digital-to-analog conversion is done at the full 100-MHz sample rate with no zero-suppression. The video output of the Tektronix scope is connected to a Video-to-IP Encoder \cite{URayVGAtoIP}.  This allows the individual signals to be monitored remotely, using VLC \cite{VLC}.  In the control room, an IP-to-Video Decoder \cite{URayIPtoVGA} is used to display the scope image continuously on a large display.
The DAQScope video stream is recorded, and the recordings are preserved for a limited time to allow delayed diagnostic of unusual detector behavior.
The settings of DAQScope (e.g. the channels being displayed) are controlled via the Run Control interface.

\subsection{Digital Sum}

The firmware allows the creation of the digital sum from all or a subset of DDC-32 channels. The first step is the summation of the waveforms of each DDC-32.  In order to ensure that the sum waveform fits into 17 bits, the three least significant bits are truncated from the sum on each DDC-32.  The sum waveform from up to 8 DDC-32s are sent to Data Sparsifiers where they are added.  The two least significant bits of the sum of eight sums is truncated to fit into 18 bits before this sum is sent to the Data Sparsifier Master. At the Data Sparsifier Master, the digital sums of the top and bottom PMTs are combined. The final digital sum has 19 bits and a noise of 1.6 LSB. 
The total digital sum is available to the Data Sparsifier Master and allows the inclusion of total area cuts in the event selection process.

\section{Performance}\label{sec:Performance}

The performance of each element of FADR has been characterized during the Quality Assurance (QA) procedures used after the production of the hardware.  Other tools have been developed to monitor the performance of FADR during regular operation.  A selection of these tools and the results of performance measurements are described in this section.

\subsection{DDC-32 Noise}

\begin{figure*}
\centering
\includegraphics[width=0.90\linewidth]{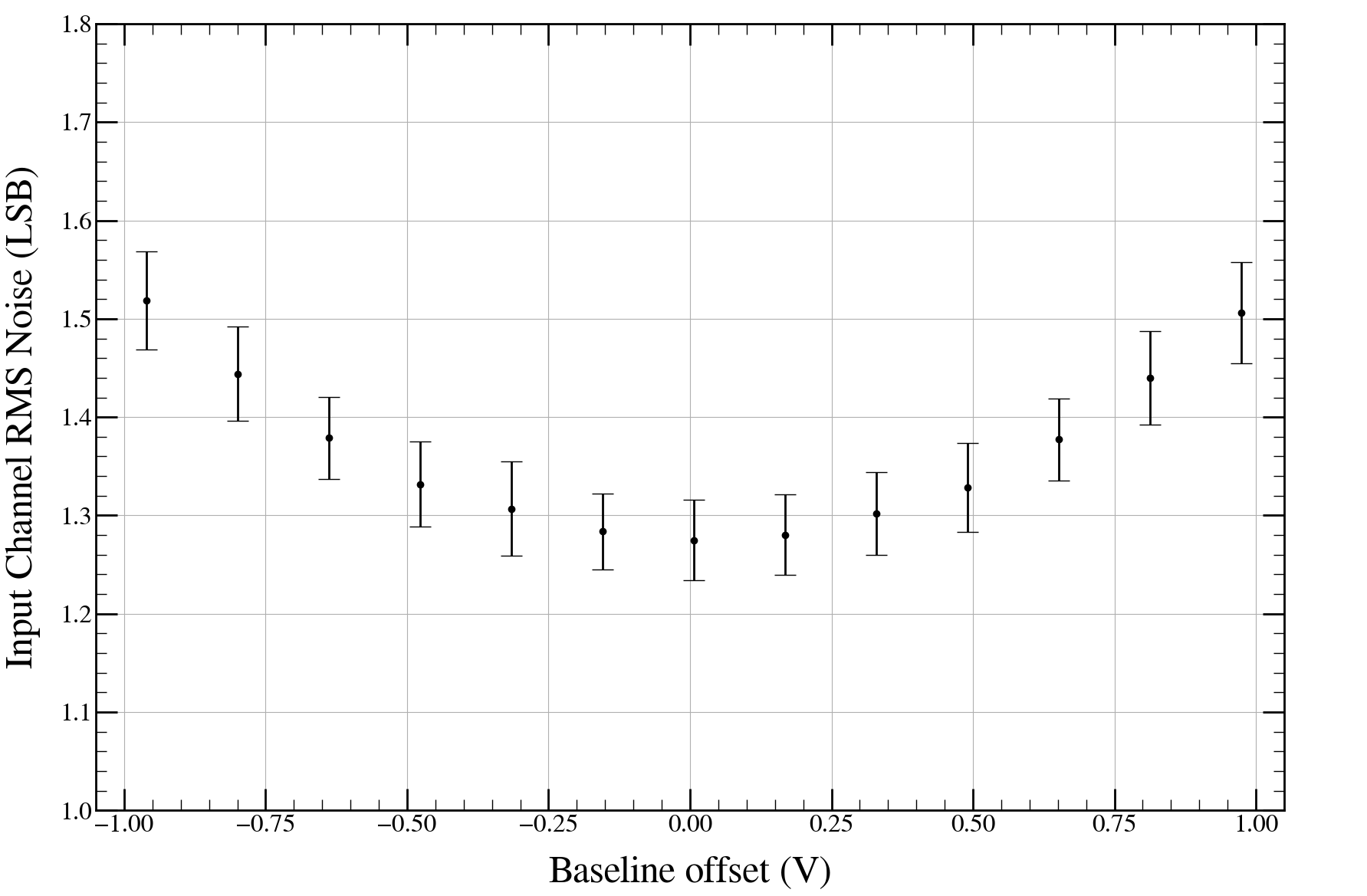}
\caption{The average measured ADC noise for all 1408 DDC-32 channels as function of the applied baseline offset.  The error bars show the standard deviation of the measured ADC noise distribution for the 1408 channels.}
\label{fig:NoiseVsBaselineOffset}
\end{figure*}

The intrinsic noise of each DDC-32 channel was measured as part of our QA procedures.  The channel noise was measured as function of the applied baseline offset.  The average channel noise for 1408 DDC-32 channels is shown in Fig.~\ref{fig:NoiseVsBaselineOffset}.  These measurements show that the lowest RMS noise is 1.3~ADCC, obtained when the baseline offset is 0~V.  When the baseline offset is changed to $\pm$1~V, the RMS baseline noise increases to 1.5~ADCC.  During normal operation, a baseline offset of +0.8~V is applied.

\subsection{DDC-32 Linearity}

\begin{figure*}
\centering
\includegraphics[width=1.00\linewidth]{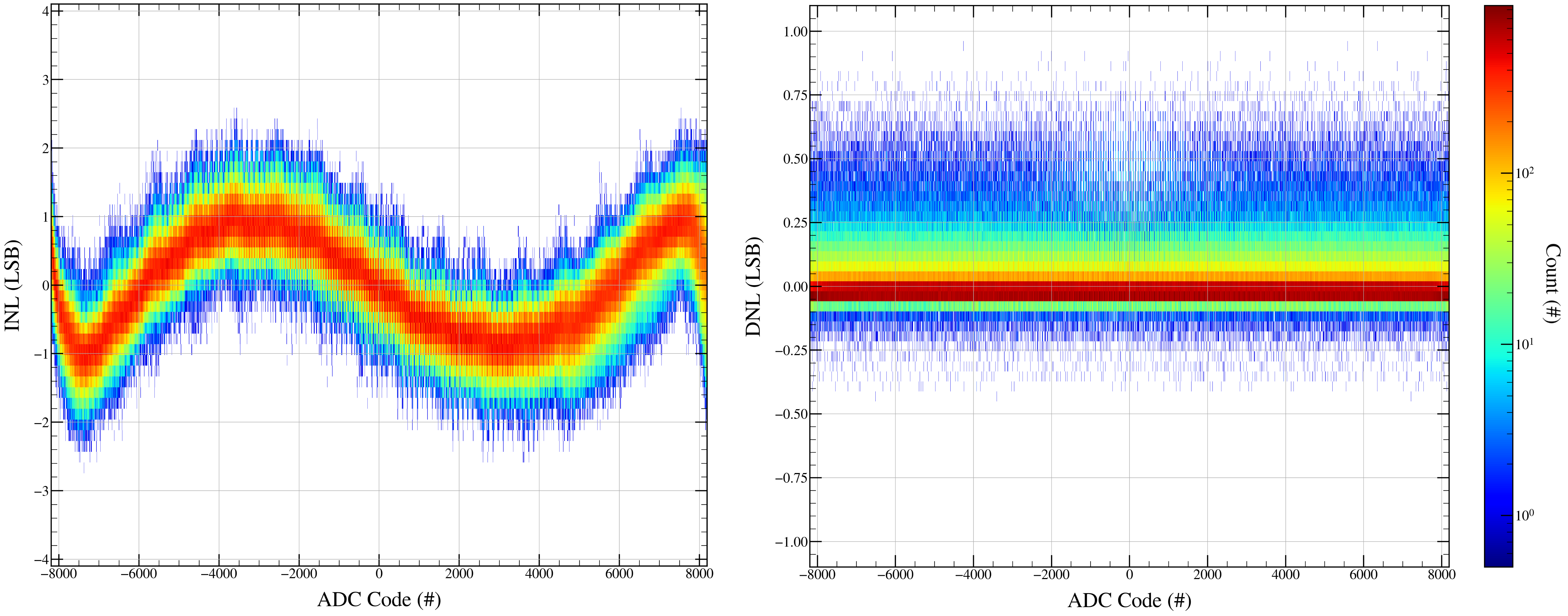}
\caption{Integral (left) and differential (right) non linearity of all 1632 DDC-32 channels.}
\label{fig:ADCLinearity}
\end{figure*}

The integral and differential nonlinearity of each DDC-32 channel was measured using a low distortion sine wave generator.  For these measurements, we used the DC1858A from Analog Devices \cite{DC1858A} which is suitable for testing ADCs with resolutions up to 18 bits.  The information collected during these measurements is preserved and can be used to correct the measured data.  The data collected for all 1632 DDC-32 channels are shown in Fig.~\ref{fig:ADCLinearity}.  The integral nonlinearity is less than $\pm$3~ADCC.  The differential nonlinearity varies between -0.5~ADCC and +1.0~ADCC.

\subsection{System Noise}\label{subsec:falseTriggerRate}

The electronic noise in the system is monitored in multiple ways.  In this section, two different approaches are discussed.

The first sequence of each run is used to extract information about the baseline noise for each channel.  The first 30 samples of each POD are used to determine the RMS baseline noise.  For all PMTs, the observed baseline noise is less than 4 ADCC.  For most PMTs, the RMS baseline noise is between 2.2 and 2.4 ADCC. This is a non-intrusive measurement.

\begin{figure*}
\centering
\includegraphics[width=0.45\linewidth]{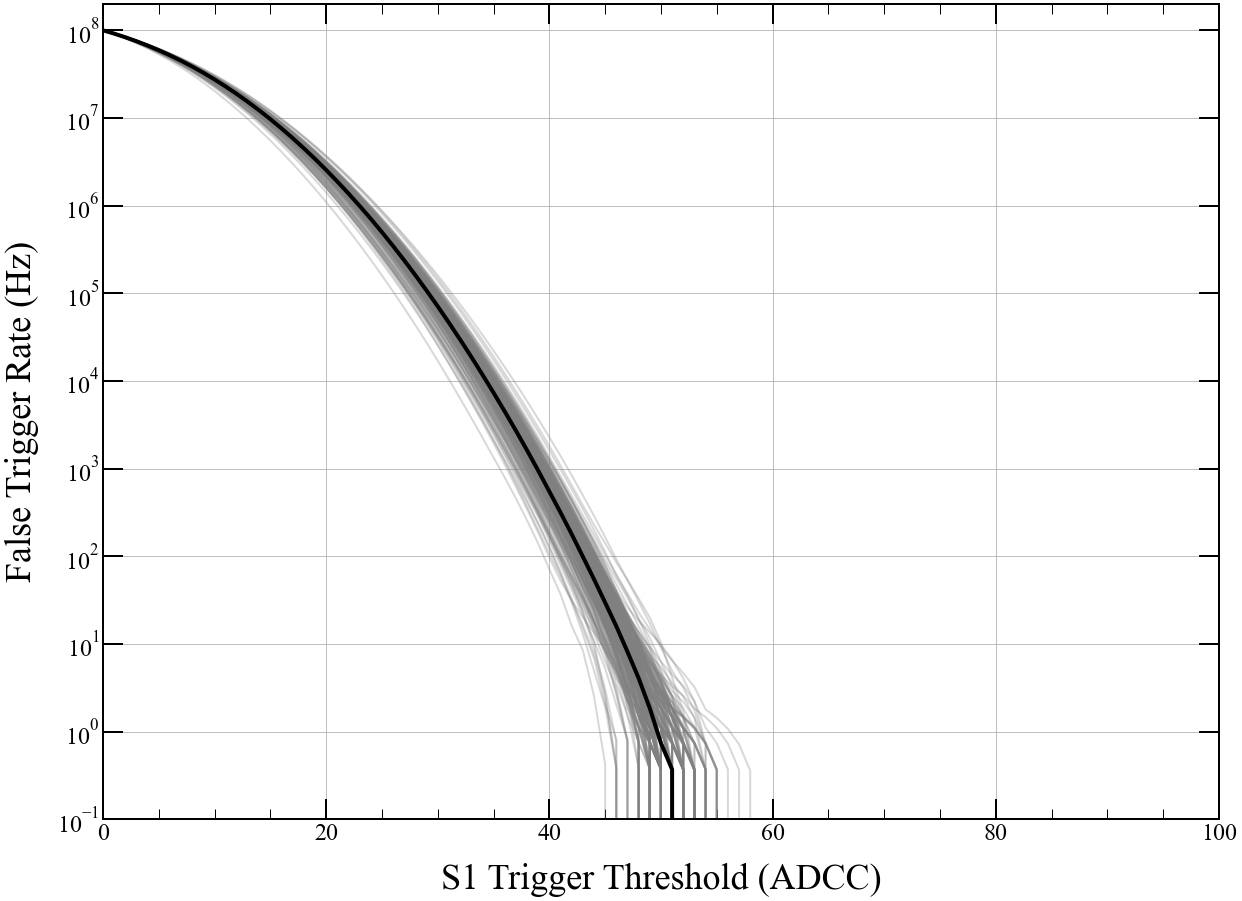}
\includegraphics[width=0.45\linewidth]{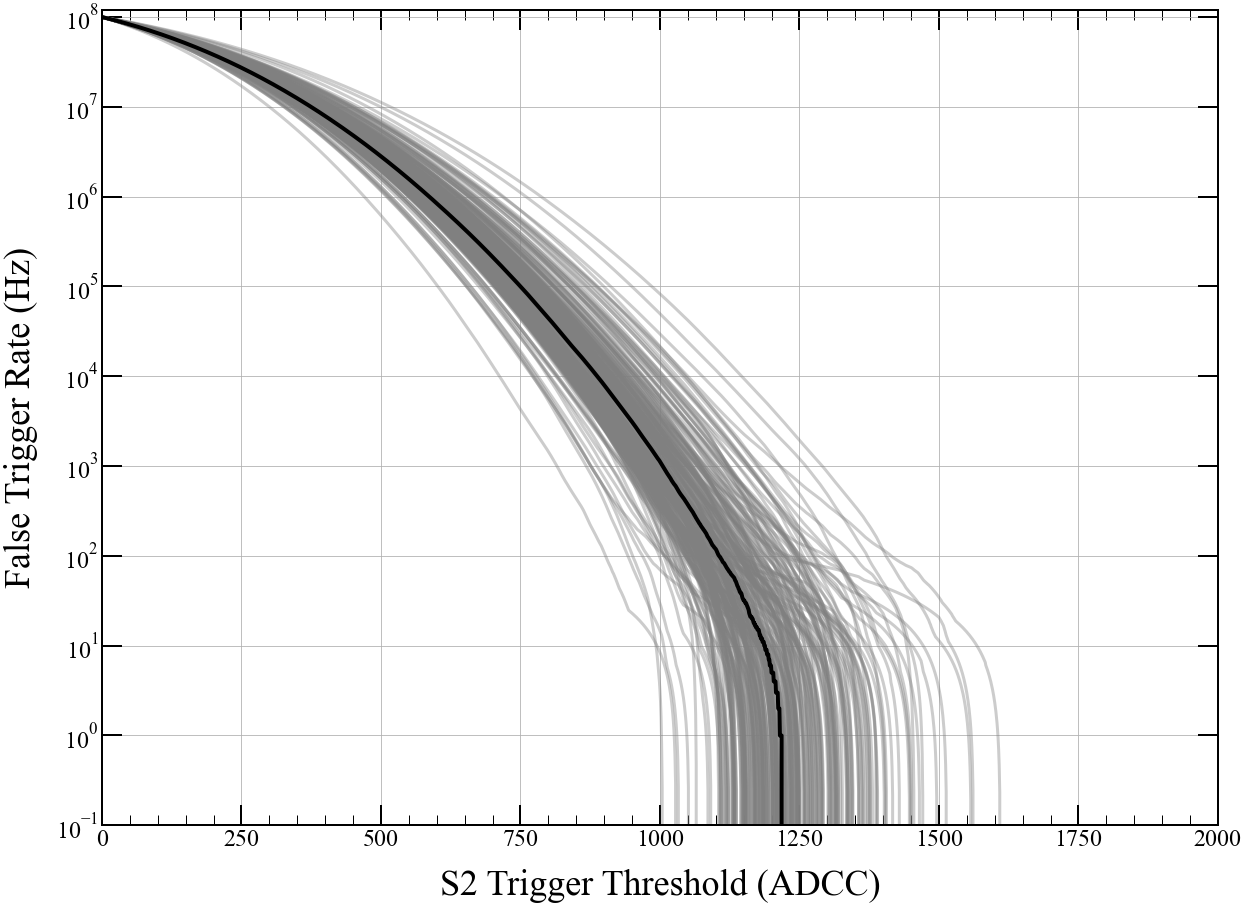}
\caption{False trigger rates for the S1 (left) and S2 (right) filters, as function of filter threshold for the top TPC PMTs.  The central lobe of the S1 filter is 5 samples wide while the central lobe of the S2 filter is 500 samples wide. The dark black curve shows the median of the distributions for the individual PMTs.}
\label{fig:falseTriggerRates}
\end{figure*}

To look for changes in noise of the electronics system, false trigger rates for both S1 and S2 filters are captured as function of filter threshold.  Examples of these distributions for a few PMTs are shown in Fig.~\ref{fig:falseTriggerRates} for a measurement carried out when the PMTs were not biased.  The filter threshold at which the trigger rate due to noise is 1~Hz is called the 1-Hz false trigger threshold and it is one of the parameters used to monitor system noise. This measurement requires a dedicated noise acquisition since it requires the high voltage of the PMTs to be off. When the PMTs are biased, the trigger rates at higher thresholds are dominated by the dark count rates in the PMTs.  The 1-Hz false trigger threshold can still be extracted by extrapolating the rate distributions for low thresholds down to 1 Hz.

The difference in the 1-Hz false trigger thresholds for the S1 and S2 filters is due to filter noise.  Since the filter calculations involve the sum over many ADC samples, the filter noise is significantly larger than the ADC sample noise.  The filter noise for the S1 filter (Eq.~\ref{eq:S1Filter}) is equal to

\begin{equation}
    \sigma_{S1}=\sqrt{\frac{3}{2}N}\sigma_{ADC}
    \label{eq:S1FilterNoise}
\end{equation}

\noindent The filter noise for the S2 filter (Eq.~\ref{eq:S2Filter}) is equal to

\begin{equation}
    \sigma_{S2}=2\sqrt{3M}\sigma_{ADC}
    \label{eq:S2FilterNoise}
\end{equation}

\noindent The ratio of the 1-Hz false-trigger thresholds shown in Fig.~\ref{fig:falseTriggerRates} is larger than the ratio of the filter noise values for the S1 and the S2 filters shown in Eqs.~\ref{eq:S1FilterNoise} and \ref{eq:S2FilterNoise}. This indicates a contribution of coherent noise to the waveforms.

\subsection{Data Transfer}\label{subsec:dataOffloading}

Each channel of the DDC-32 has two circular buffers.  Data can be written to one buffer while data from the other buffer is being transferred. For each buffer, the start and end times are captured in the data stream. The buffer start time is the time that a buffer is ready to receive data; the buffer end time is the time of the end of the event or when the first buffer of a particular digitizer is filled.  Different digitizers can have different buffer start and end times.  If the buffer stop time of event $n$ is larger than the buffer start time of event $n+1$ there is no dead time associated with switching buffers.  If the buffer stop time of event $n$ is less than the buffer start time of event $n+1$ there is period when no data could be captured and dead time is incurred. This happens when one buffer is transferring data and the second buffer is waiting to transfer data.  To offload one full buffer (18,412 samples) requires 340~$\mu$s.  The maximum data transfer time when all 32 channels have a full buffer is 11~ms.  The data from the 14 Data Extractors are transferred to 14 Data Collectors.  Currently we can sustainably acquire data at 1200~MB/s.  

The average time to process data files on the Data Collectors and create event files on the Event Builders is 22~s. The average time to transfer the event files from the Event Builders to the RAID arrays on the surface is 4~s. If writing to disk fails, our monitoring system will raise an alarm.

To ensure data integrity we have implemented multiple levels of checksumming. First, every byte of data for each event is checksummed (CRC32) as it is read out from the Block RAMs of the DDC-32 FPGAs. The Data Extractors, as they split individual event data into jumbo frame-sized chunks with a maximum size of 8,800 bytes, checksum each UDP packet payload that is sent to the Data Collectors. The Data Collectors cross-check the payload checksums of each received UDP packet, and perform an event-level checksum cross-check once a given event is stitched back together. All the server machines in the data pipeline (Data Collectors, Event Builders, etc) use ECC-enabled memory modules and all file systems in the data chain use data and metadata checksumming. As ROOT sequence files are being generated on the Event Builders, their SHA256 checksums are calculated and stored in a data-tracking DB. These SHA256 checksums are recalculated and cross-checked at NERSC, before a sequence file is marked as successfully transferred and permanently stored.

\subsection{Livetime}\label{subsec:daqLiveTime}

\begin{figure*}
\centering
\includegraphics[width=0.9\linewidth]{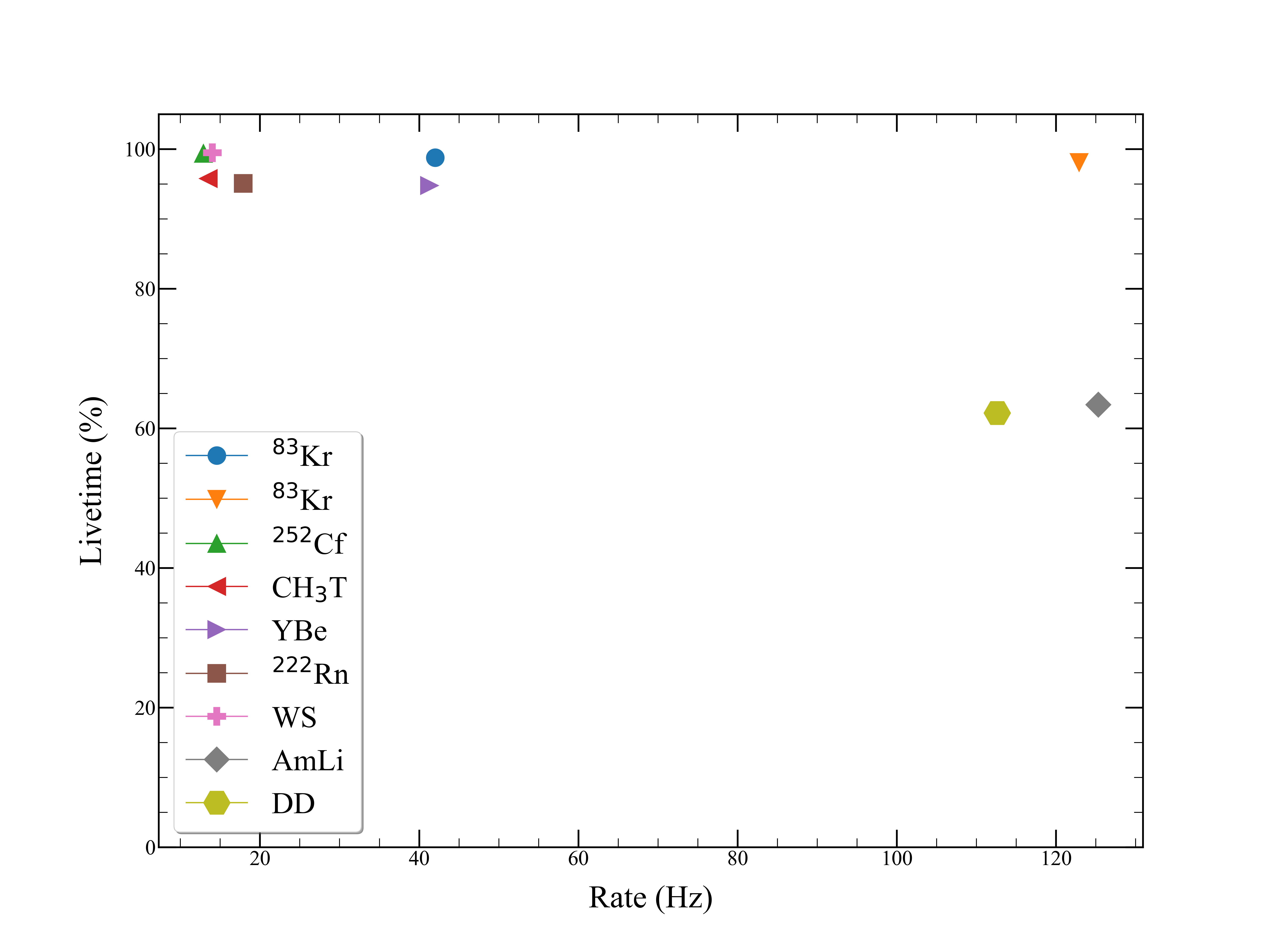}
\caption{Measured FADR livetime during WIMP search and various calibration measurements as function of event rate. The livetime shown is the raw DAQ livetime and reflects the fraction of time the DAQ is able to accept triggers.  The livetime is not only a function of event rate, but also event size.  This is clearly shown by the measurements with the AmLi source and DD neutrons. For these sources, the event size is three times larger than the event size during WIMP Search (WS). Measurements with $^{83m}$Kr for which the event size is similar to the WS event size obtained a 98\% DAQ livetime at a rate of 123 Hz.}
\label{fig:daqLiveTime}
\end{figure*}

Using the buffer start and stop times discussed in Sec.~\ref{subsec:dataOffloading}, a detailed calculation of the livetime can be carried out.  The buffer information can be used to determine the exact times at which the DDC-32s are unable to accept triggers and thus the livetime. An example of the measured livetimes during WIMP Search and detector calibrations is shown in Fig.~\ref{fig:daqLiveTime} as function of event rate.  The livetime is also a function of event size.  This is clearly visible in Fig.~\ref{fig:daqLiveTime} which shows a 98\% livetime obtained for $^{83m}$Kr calibrations at 122 Hz while the livetimes obtained during AmLi and DD neutron calibrations is around 63\% at similar event rates. The event sizes for AmLi and DD neutron calibrations are about 3 times as large as the event sizes during $^{83m}$Kr calibrations.

During WIMP Search, the livetime due to filled buffers is 99.5 $\pm$ 0.2\%.  With a trigger holdoff time of 2 ms, the livetime is reduced to 95.9 $\pm$ 0.4\%.

FADR also provides information about the inability to write PODs to the circular buffers during periods of high detector activity. Under these circumstances, the buffers of some channels may become filled, preventing complete PODs from being written to these buffers.  If this happens, the PODs are flagged as truncated PODs.  During offline analysis, decisions can be made on how to handle this type of events.

\subsection{Event Selection}\label{subsec:eventSelection}

To verify proper event selection, the properties of captured events are compared to the configuration settings of FADR.  Since it is known what trigger caused the event capture, the raw waveforms can be processed by the corresponding S1 or S2 filter, and the reconstructed PMT multiplicity can be compared with the multiplicity requirement used for that trigger.  As an example, the S1 trigger for the OD will be discussed.

During SR1, the OD waveforms are processed by an S1 filter with the following parameters:
\begin{enumerate}
    \item Filter threshold = 3,000 ADCC
    \item Filter width = 15 samples (central lobe)
    \item Coincidence window = 32 samples
    \item Minimum required multiplicity = 15
\end{enumerate}

\begin{figure*}
\centering
\includegraphics[width=1.0\linewidth]{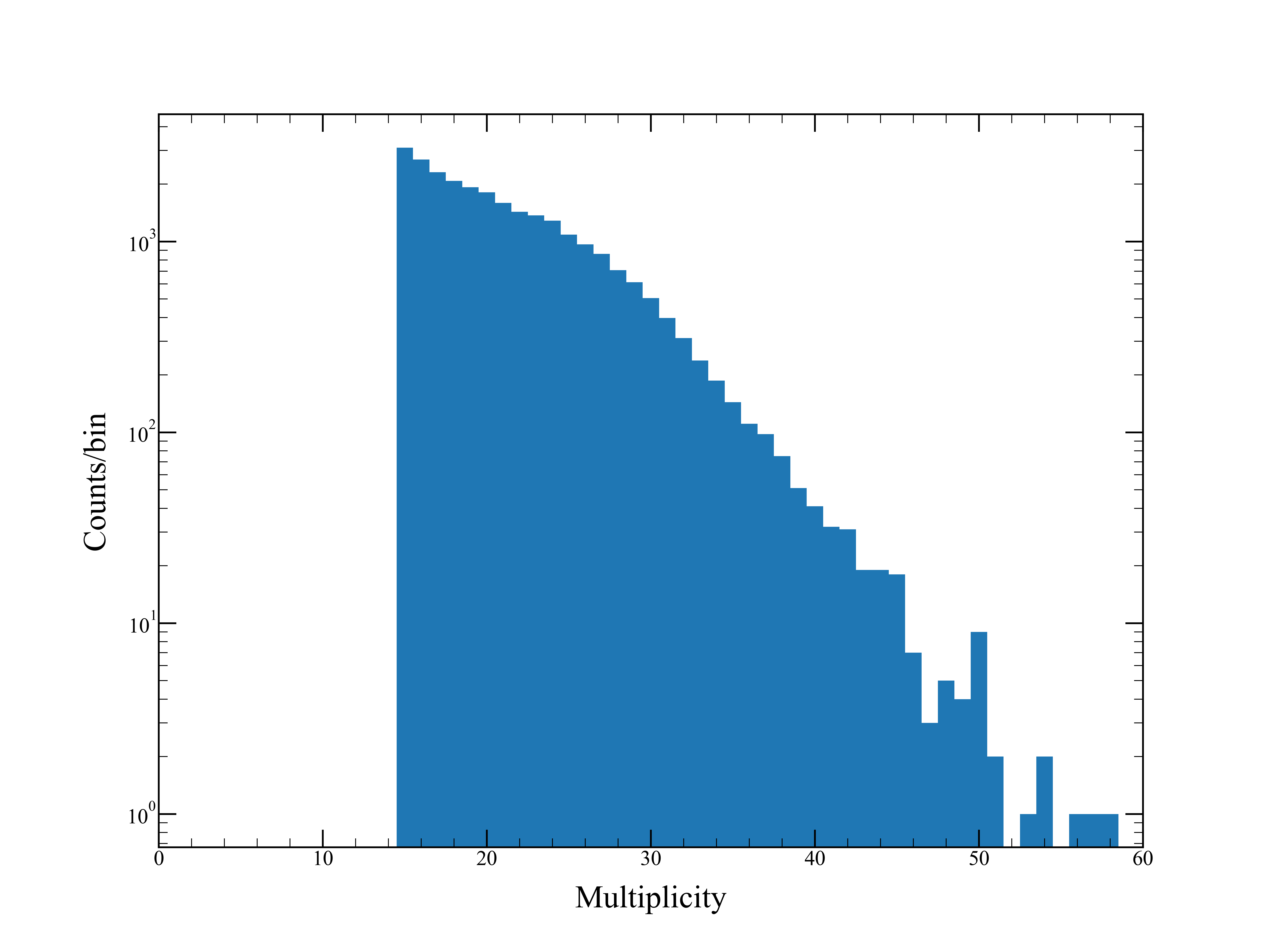}
\caption{OD multiplicity of events associated with a valid OD S1 trigger.  The S1 trigger required a threshold crossing of the S1 filter of at least 15 PMTs during a 320 ns coincidence window. }
\label{fig:ODS1Mult}
\end{figure*}

\noindent For events triggered by the OD trigger, the captured PODs were processed with an S1 filter and the multiplicity of threshold crossings within the coincidence window was determined.  A histogram of the reconstructed multiplicity is shown in Fig.~\ref{fig:ODS1Mult}.  None of the 26,174 events examined have a multiplicity less than 15.  The multiplicity requirement is met about 93 samples (930 ns) before the trigger time stamp of the event, indicating the latency involved with the propagation of the S1 filter signals in the Data Sparsifier chain.

Similar studies are also carried out for the S2 filters.

\subsection{Zero suppression}\label{subsec:zeroSuppression}

Since only PODs are preserved, a thorough study of the POD algorithm implemented on the FPGA was carried out to verify that no loss of information occurs when PODs are extracted from the raw waveforms.

\begin{figure*}
\centering
\includegraphics[width=1.0\linewidth]{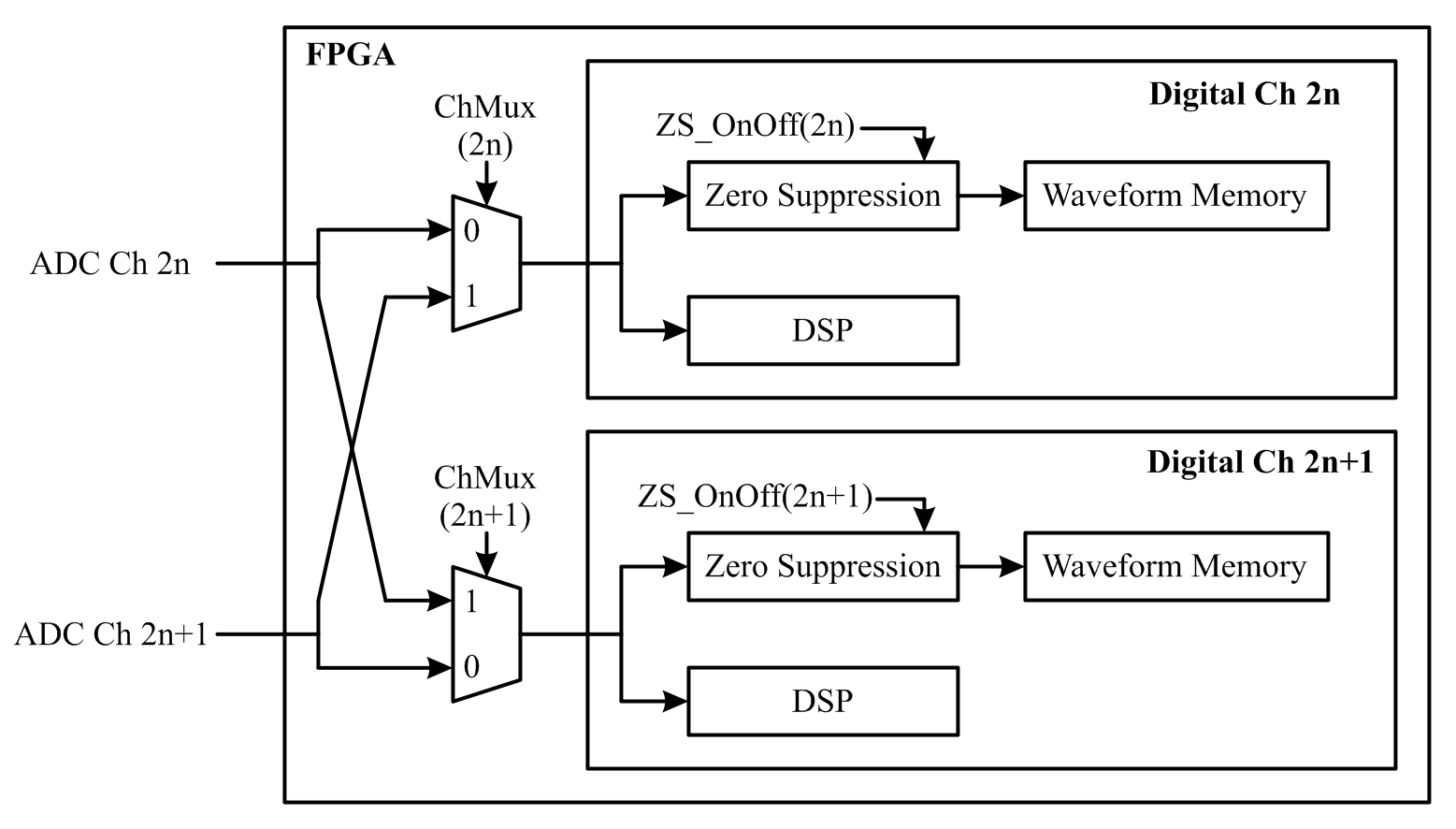}
\caption{Simultaneous data collection of raw and POD data. In this mode, the output of even or odd ADCs are directed towards adjacent FPGA channels. For example, if ChMux10=0, ChMux11=1, ZS\textunderscore OnOff10=On, ZS\textunderscore OnOff11=Off, then the ADC Channel 10 data will simultaneously be recorded in zero-suppressed form in the waveform memory of Digital Channel 10 and raw format in the waveform memory of Digital Channel 11.}
\label{fig:ZAandRaw}
\end{figure*}

The proper operation of the POD algorithm was verified by collecting LED data in a special mode, where the waveform from specific ADC channels are directed to two FPGA channels.  This mode is shown schematically in Fig.~\ref{fig:ZAandRaw}. One FPGA channel captures the 500 samples of the waveform after the trigger signal while the other channel operates in POD mode and only captures data when the POD threshold is crossed.  Such data sets allow us to determine the efficiency of the POD algorithm by performing the studies that are described in detail in this section.  

\begin{figure*}
\centering
\includegraphics[width=1.0\linewidth]{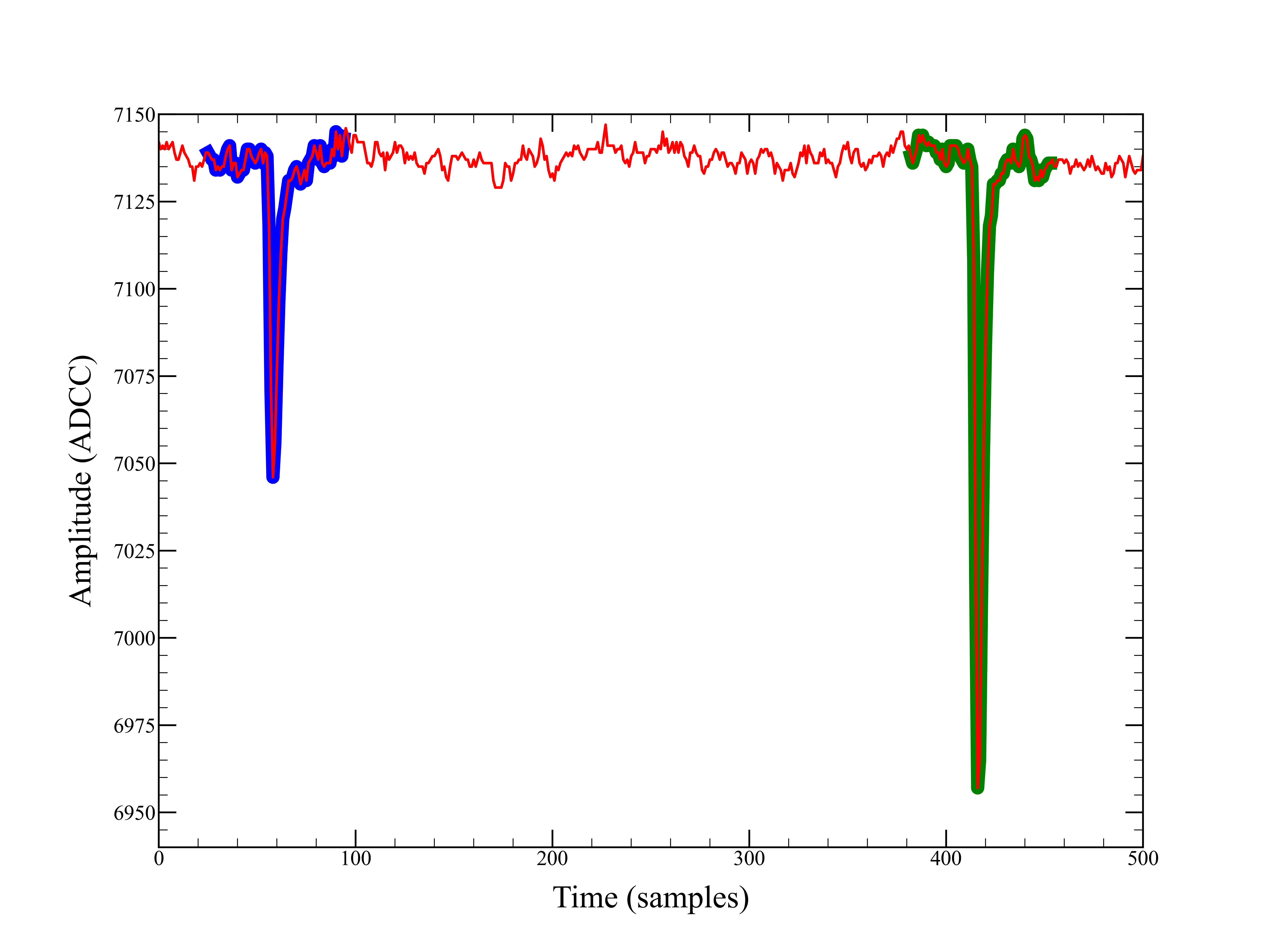}
\caption{Example of a raw waveform with the corresponding PODs that were extracted by the FPGA.  In this example, two PODs were identified.}
\label{fig:LZPods}
\end{figure*}

\subsubsection{Bit and Timestamp Consistency}\label{subsubsec:bitConsistency}
To determine the bit and timestamp consistency of the captured PODs, the samples in each POD were compared to the corresponding samples in the raw waveforms.  An example is shown in Fig.~\ref{fig:LZPods} where two PODs were captured by the FPGA.  In this Figure, the PODs are overlayed with the raw waveform. showing a perfect match, both in time and in amplitude.   A total of about $6 \times 10^7$ PODs, equivalent to about $4 \times 10^9$ ADC samples, were compared with the corresponding raw waveforms.  No inconsistencies between the sample values in the PODs and in the raw waveforms were observed.

\subsubsection{POD Consistency}\label{subsubsec:podConsistency}
Two additional studies were carried out to verify the proper operation of the POD algorithm.  We simulated the FPGA code to compare POD crossings observed in the raw waveforms to the PODs captured by the FPGA.  To ensure that the POD algorithm did not miss any peaks, we compared the PODs captured by the FPGA with peaks identified in the raw waveforms by a peak finding algorithm that is part of the python package SciPy \cite{SciPy}. These studies used data files collected in the dual-capture mode discussed above with a POD threshold of 25 ADCC.

\subsubsection{Simulate the FPGA code.}

In this analysis, the code that is used on the FPGA to determine if a POD crossing is observed in the raw waveform is simulated in python.  If a POD crossing is observed, we determine if the corresponding POD is also captured.  0.00007\% of the POD crossings in the raw waveforms are observed to have no corresponding POD, all of which have an amplitude within 1 ADCC of the POD threshold. No POD inconsistencies were observed for amplitudes greater than 1 ADCC above the POD threshold.

\subsubsection{Studies using scipy.signal.find\textunderscore peaks.}

In this analysis, scipy.signal.find\textunderscore peaks \cite{SciPyFindPeaks} is used to identify peaks in the raw waveforms.  The peaks found are then compared to the captured PODs.  All of the observed inconsistencies are due to differences in the way the FPGA and scipy.signal.find\textunderscore peaks define the baseline. On the FPGA, the baseline is defined by the previous 32 samples; in scipy.signal.find\textunderscore peaks, the  baseline is defined by the local minimum of the samples surrounding an identified peak.  The deviation from the baseline of peaks found by scipy.signal.find\textunderscore peaks may be larger than 25 ADCC, but this larger value is entirely due to differences in the estimated baseline.  In the $4 \times 10^7$ PODs studied, no inconsistencies were found between peaks identified by scipy.signal.find\textunderscore peaks in the raw waveforms and the captured PODs that were not due to differences in baseline evaluation.

\section{Summary}\label{sec:Summary}

The design of the data acquisition system for LZ, FADR, was based on our experience with LUX and the required calibration rates and data volume for LZ. The RMS baseline noise of the entire signal processing chain for most PMTs is between 2.2 and 2.4~ADCC ($\approx$0.3~mV), ensuring a high efficiency for the detection of single photoelectrons with amplitudes of around 50~ADCC (6~mV). In order to reduce the impact of slow baseline variations, the PMT waveforms are processed with digital filters that perform automatic baseline subtraction and waveform integration before event selection.  The design of FADR is optimized for data transfer. The data for a specific event are written to 14 Data Collectors.  The overall performance of the system is limited by the Data Collector that sees the highest data volume.  Currently we can sustainably acquire data at 1200~MB/s.

The hardware and firmware of FADR were developed by SkuTek \cite{SkuTek} and the University of Rochester. The system is designed around the Kintex-7 FPGA. FADR digitizes 1359 PMT channels with 14-bit, 100-MHz, 2-V ADCs. The waveforms are processed by digital filters that are sensitive to S1 and S2 signals. Event selections are made using the information provided by these digital filters.

Pulse Only Digitization reduces the raw waveform data volume by a factor of up to 50. The PODs extracted from the raw waveforms are stored in circular buffers. Once an event of interest has been defined, the PODs that fall within the event window are preserved and written to disk.  Several detailed studies were carried out to verify that no loss of information occurs when PODs are extracted from the raw waveforms.

Using the buffer start and stop times for each PMT, a detailed calculation of the FADR livetime can be carried out. During WIMP Search, the livetime due to filled buffers is 99.5 $\pm$ 0.2\%.  With a trigger holdoff time of 2 ms, the livetime is reduced to 95.9 $\pm$ 0.4\%.

Since we are in full control of the firmware of FADR, we developed additional features that enhance our ability to test, troubleshoot, monitor, and expand the information on which event selection is based.  The flexibility provided by the FPGAs allows us to monitor the performance of the detector and FADR in parallel to normal data acquisition. Examples were presented.

\section{Acknowledgements}

The research supporting this work took place in whole or in part at the Sanford Underground Research Facility (SURF) in Lead, South Dakota. Funding for this work is supported by the U.S. Department of Energy, Office of Science, Office of High Energy Physics under Contract Numbers 
DE-AC02-05CH11231, 
DE-SC0020216, 
DE-SC0012704, 
DE-SC0010010, 
DE-AC02-07CH11359, 
DE-SC0015910, 
DE-SC0015910, 
DE-SC0014223, 
DE-SC0010813, 
DE-SC0019193, 
DE-SC0009999, 
DE-NA0003180, 
DE-SC0011702, 
DE-SC0010072, 
DE-SC0015708, 
DE-SC0006605, 
DE-SC0008475, 
DE-FG02-10ER46709, 
UW PRJ82AJ, 
DE-SC0013542, 
DE-AC02-76SF00515, 
DE-SC0018982, 
DE-SC0019066, 
DE-SC0015535, 
DE-SC0019319, 
DE-AC52-07NA27344, 
DE-SC0024114, and 
V\& DOE-SC0012447.	
SkuTek Instrumentation is partly supported under the DOE SBIR grant DE-SC0009543.  This research was also supported by U.S. National Science Foundation (NSF); the UKRI’s Science \& Technology Facilities Council under award numbers 
ST/M003744/1, 
ST/M003655/1, 
ST/M003639/1, 
ST/M003604/1, 
ST/M003779/1, 
ST/M003469/1, 
ST/M003981/1, 
ST/N000250/1, 
ST/N000269/1, 
ST/N000242/1, 
ST/N000331/1, 
ST/N000447/1, 
ST/N000277/1, 
ST/N000285/1, 
ST/S000801/1, 
ST/S000828/1, 
ST/S000739/1, 
ST/S000879/1, 
ST/S000933/1, 
ST/S000844/1, 
ST/S000747/1, 
ST/S000666/1, 
ST/R003181/1; 
Portuguese Foundation for Science and Technology (FCT) under award numbers PTDC/FIS-PAR/2831/2020; the Institute for Basic Science, Korea (budget numbers IBS-R016-D1). We acknowledge additional support from the STFC Boulby Underground Laboratory in the U.K., the GridPP \cite{Faulkner_2006,Britton_2009} \cite{Britton_2009} and IRIS Collaborations, in particular at Imperial College London and additional support by the University College London (UCL) Cosmoparticle Initiative. The LZ Collaboration acknowledges key contributions of Dr. Sidney Cahn, Yale University, in the production of calibration sources. This research used resources of the National Energy Research Scientific Computing Center, a DOE Office of Science User Facility supported by the Office of Science of the U.S. Department of Energy under Contract No. DE-AC02-05CH11231. The University of Edinburgh is a charitable body, registered in Scotland, with the registration number SC005336. The assistance of SURF and its personnel in providing physical access and general logistical and technical support is acknowledged.

\bibliography{main}

\end{document}